\documentclass[12pt,preprint]{aastex}
\usepackage{epsf}
\usepackage{emulateapj5}
\usepackage{onecolfloat}
\usepackage{apjfonts}
\usepackage{amsmath}
\def\BibTeX{{\rm B\kern-.05em{\sc i\kern-.025em b}\kern-.08em
    T\kern-.1667em\lower.7ex\hbox{E}\kern-.125emX}}

\newcommand{\ltsim}{\lower0.6ex\vbox{\hbox{$ \buildrel{\textstyle <}\over{\sim}\ $}}}
\newcommand{\gtsim}{\lower0.6ex\vbox{\hbox{$ \buildrel{\textstyle >}\over{\sim}\ $}}}

\newcommand{\beq}{\begin{equation}}
\newcommand{\eeq}{\end{equation}}

\def\fmerger{f_{\rm merger}}
\def\fmin{f_{\rm min}}
\def\fqnr{f_{\rm qnr}}
\def\zre{z_{\rm re}}
\def\tmig{t_{\rm mig}}
\def\epsilonm{\epsilon_{\rm merger}}
\def\epsilonr{\epsilon_{\rm ringdown}}

\newcommand{\tfs}{t_{\mathrm{fs}}}

\newcommand{\hMsun}{\ h^{-1}\mathrm{M}_{\odot}}

\newcommand{\Msun}{M_{\odot}}
\newcommand{\kms}{{\,{\rm km}\,{\rm s}^{-1}}}
\newcommand{\kpc}{{\,{\rm kpc}}}

\newcommand{\si}{{\rm{s}^{-1}}}

\newcommand{\Omegagw}{\Omega_{\mathrm{gw}}}
\newcommand{\Egw}{\overline{E}_{\mathrm{gw}}}
\newcommand{\fcb}{f_{\mathrm{cb}}}

\newcommand{\rhocrit}{\rho_{\mathrm{c}}}

\newcommand{\Vesc}{V_{\mathrm{esc}}}
\newcommand{\Vmax}{V_{\mathrm{max}}}

\newcommand{\Rvir}{R_{\mathrm{vir}}}
\newcommand{\Mvir}{M_{\mathrm{vir}}}

\newcommand{\rs}{r_{\mathrm{s}}}

\newcommand{\mbh}{M_{\mathrm{BH}}}

\newcommand{\ms}{M_{\mathrm{S}}}
\newcommand{\zbh}{z_{\mathrm{BH}}}
\newcommand{\svd}{\sigma_{\star}}
\newcommand{\Mcrit}{M_{\mathrm{v,crit}}}
\newcommand{\rpp}{r_{\mathrm{pp}}}
\newcommand{\vrecoil}{V_{\mathrm{recoil}}}

\newcommand{\SN}{\langle {\rm S} / {\rm N} \rangle }

\newcommand{\dd}{\mathrm{d}}

\font\FermiSmallfont=cmssq8 scaled 1200

\def\LANLppthead#1#2{
\null
\begin{center}\vskip -1.0truein{\hbox to 7.5truein {
\vbox to 1.1in {\vfill \FermiSmallfont
               \vfill}
\hfill
\vbox to 1.1in {\vfill \FermiSmallfont
               \hbox{#1}
               \hbox{#2}
               \vfill}
}}\vskip-0.0truein\end{center}}

\begin{document}
\LANLppthead {LA-UR-05-8246}{astro-ph/0503511}

\vspace{1mm}
\slugcomment{{\em The Astrophysical Journal, submitted}}

\shortauthors{KOUSHIAPPAS \& ZENTNER}

\title{Testing models of supermassive black hole seed formation through 
gravity waves}

\author{Savvas M. Koushiappas,\altaffilmark{1,2}
  Andrew R. Zentner,\altaffilmark{3}
  }
\vspace{2mm}

\begin{abstract}

We study the gravitational wave background produced from the formation 
and assembly of supermassive black holes within the cosmological 
paradigm of hierarchical structure formation.  
In particular, we focus on a supermassive black hole 
formation scenario in which the present-day population of 
supermassive black holes is built from 
high-mass seed black holes and we compute the 
concomitant spectrum of gravitational radiation produced 
by mergers of the seed black holes.  
We find that this scenario predicts a large, 
gravitational wave background that should be 
resolved into individual sources 
with space interferometers such as the proposed 
{\em Laser Interferometric Space Antenna} (LISA).  The number of 
inspiral, merger and ringdown events above a signal to noise 
ratio of 5 that result from massive black 
hole seeds is of order $10^3$. This prediction 
is robust and insensitive to several of the details of 
the model.  We conclude that an interferometer such as LISA will be able to 
effectively rule out or confirm a class of models where 
supermassive black holes grow from high-mass seed black holes 
and may be able to place strong limits on the role of mergers 
as a channel for supermassive black hole growth.  Supermassive black 
hole seeds likely form in the earliest proto-galactic structures 
at high redshift and the masses of supermassive black holes 
are known to be strongly correlated with the potentials of the 
spheroids in which they reside, therefore these results imply that 
space interferometers can be used as a powerful probe of the physics of 
galaxy formation and proto-galaxy formation at very high 
redshift.

\end{abstract}


\altaffiltext{1}{Department of Physics, ETH-Z\"urich, CH-8093 Z\"urich, Switzerland}
\altaffiltext{2}{T-6 Theoretical Division \& ISR-1 ISR Division, The University of California, Los Alamos National Laboratory, NM 87545, USA; smkoush@lanl.gov}

\altaffiltext{3}{Kavli Institute for Cosmological Physics 
       and Department of Astronomy and Astrophysics, 
       The University of Chicago, Chicago, IL 60637,USA; zentner@kicp.uchicago.edu}

\section{INTRODUCTION}
\label{section:introduction}

The existence of supermassive black holes (SBHs) in the centers 
of galaxies has been the subject of numerous recent studies. 
Stellar and gas-dynamical measurements in nearby galaxies and 
reverberation mapping on high-redshift active galactic nuclei 
(AGN) have shown that SBHs with masses 
$\mbh \sim 10^6-10^9 \, \Msun$ reside 
in the centers of spheroidal systems, 
these being elliptical galaxies and 
the bulges of spiral galaxies 
\citep[for a review, see][]{F04}.
The inferred black hole masses together with measurements
of the absolute luminosity of the host spheroids, have 
revealed a correlation between the mass of the central supermassive 
black hole, $\mbh$, and the mass of its host spheroid $\ms$ 
\citep{KR95, M98, MF01a}.
Similarly, detailed studies of the stellar velocity 
dispersions $\sigma_{\star}$, in the host spheroids 
have revealed an even tighter correlation with the 
mass of the black hole that it is harboring 
\citep{FM00, MF01b, TETAL02, GETAL00}.  
The black hole (BH) and the host spheroid 
``know'' of the presence of one another, 
though the explanation(s) of this correlation is(are) 
unknown. 

Models of SBH formation and growth should 
address the following questions.  
\begin{itemize}

\item[1.] When and where do supermassive black holes, and/or 
the {\em seed} black holes from which they grow, form?

\item[2.] How do these black holes evolve 
to the masses that we observe in the local universe?

\item[3.] What is the fundamental coupling that 
connects the stellar dynamics of the bulge to 
the central black hole?

\item[4.] What testable predictions can be used 
to falsify the model?

\end{itemize}
In this paper, we show how gravitational wave detectors 
can be used to test models of seed BH formation. 
Models in which SBHs form from high-mass ($\mbh \gg 10^3 \Msun$) 
seeds can be tested explicitly with upcoming gravity wave experiments.  
As a particular example, we show that the SBH seed formation 
scenario proposed by \citet[][henceforth KBD]{KBD04} makes 
the robust prediction of a large 
gravity wave background that 
should be measurable with a space interferometer, such as the 
{\em Laser Interferometric Space Antenna} 
(LISA)\footnote{\tt http://lisa.nasa.gov/}, now in its 
formulation stage and planned for launch in 2015.  

This is a result with interesting consequences.  
It provides a method for falsifying models with 
high-mass black holes seeds, such as the KBD model, 
for studying the parameters of the KBD-like models, 
for assessing the importance of mergers in the formation of 
SBHs, and the establishment of the observed correlations 
between SBHs and their host spheroids. 
As we elaborate on below, an important implication then 
is that LISA can be used as a tool to study the 
physics of galaxy formation.  
Moreover, if it were realized in Nature, 
the gravity wave background that we compute, 
would represent a significant background 
contaminant to instruments like LISA and 
any others that are envisaged for operation 
in the low-frequency regime, such as the 
Decihertz Interferometer Gravitational Wave Observatory 
\citep[DECIGO,][]{seto_etal01} or the 
Big Bang Observer 
(BBO)\footnote{\tt http://universe.nasa.gov/program/bbo.html}, 
which could otherwise aim at 
measuring gravity wave backgrounds arising 
from the epoch of inflation 
\citep[e.g.][]{turner97}, 
as a consequence of theories with large extra 
dimensions \citep{hogan00}, or from 
supernovae \citep{buonanno_etal05}.

Numerous scenarios have been proposed for the formation of 
SBHs.  These models address the first three of the above 
questions through the dynamical evolution of a stellar system 
\citep[e.g., ][]{QS90, HB95, L95, EETAL01, MEK03, AGR03, AGR01}
or through the hydrodynamical evolution of a pressure-supported 
object \citep[e.g., ][]{HR93, LR94, EL95, HNR98, G01, BL03, KBD04}.  
In particular, \citet{MR01} (MR01 hereafter) proposed a model 
where the seeds of SBHs formed at $z \sim 20$ in rare 
halos ($\sim 3\sigma$ peaks in the smoothed density field) 
that are sufficiently massive that molecular hydrogen 
cooling allows runaway collapse of a significant amount of 
baryons.  These baryons form a pressure-supported object 
of mass $\sim 260 \Msun$ that is unstable and collapses 
to a BH on a timescale of $\sim [1-10]$ Myr.  
Seed BHs in this model have a typical mass 
$\mbh \sim [1-2] \times 10^2 \Msun$.  This model 
is motivated by numerical studies that simulate an 
extremely dense environment at high-redshift in an attempt to 
identify the time and environment where the very first luminous 
objects in the universe form \citep{ABN00, BCL99} and evolve 
\citep{FWH01, BHW01, HW02, SS02, S02}. 
The ramifications of such a scenario were 
investigated in detail by \citet{VHM03} who demonstrated 
that subsequent mergers experienced by seed BHs during the 
formation of a galactic halo do not play a 
significant role in the growth of the SBH.  Instead, 
a finely-tuned accretion mechanism ensures that SBHs 
grow in such a way that the relationship 
between $\mbh$ and $\svd$ is preserved.

Alternatively, KBD proposed a model where 
seed BHs form from the low angular momentum 
material in proto-galactic disks at high redshift.  
This model predicts seed BHs of a specific mass scale, 
$\mbh \sim \mathrm{a\ few} \times 10^5 \Msun$ that reside 
in rare, massive halos at high redshift ($z \gtsim 12$).  
BH growth via mergers is more important 
in this scenario and KBD demonstrated that 
the the slope of the relationship between 
$\mbh$ and $\ms$ can be {\em set by hierarchical merging}, 
while the normalization of this relationship may be due 
to an accretion epoch at some lower redshift. 
Additionally, KBD showed that if central BHs 
and their host spheroids are built by mergers, 
galaxies that do not contain BHs, will also be bulge-less.  
The transition is set at approximately 
$\ms \sim {\rm few} \times 10^9$ and 
$\mbh \sim 10^5$, and naturally explains the 
observational indication of a lower limit on $\mbh$. 

Unfortunately, it is difficult to predict robustly 
observational signatures that can constrain or distinguish 
between SBH formation scenarios.  SBHs may 
be assembled primarily through mergers of 
numerous small BHs or relatively fewer 
mergers of massive BHs. It may be the case 
that SBHs are formed through periods of possibly 
super-Eddington accretion \citep{KETAL04}. 
One promising avenue for shedding light on the 
SBH formation mystery is the possible detection 
of the gravitational waves emitted from mergers during 
their hierarchical assembly \citep{TB76}.  A space-based, 
interferometric gravity wave detector, such as the 
proposed LISA interferometer, should 
be capable of detecting such merger events 
\citep{T95,thorne96,FH98,flanagan_hughes98b,cornish_larson01}.  
For such an instrument, mergers of BHs with 
masses $\mbh \gtsim 10^3 \Msun$ at redshifts from 
$z \sim 0-30$ will be observable as strain perturbations, 
providing a new test of the viability of SBH 
growth through mergers.

The relative importance of mergers and accretion in 
the growth of SBHs is still under debate. 
Even if accretion is the dominant growth mechanism, 
as suggested by studies aiming at explaining  
the quasar luminosity function 
\citep{GLETAL99,SSMD99,SW03,HM00,HMSMS03} and 
the effect of feedback in the central regions of 
galaxies and clusters \citep{PS04,MQT04,DB04,SOCS04}, 
the paradigm of hierarchical structure formation 
predicts that some growth must be due to mergers.  
In fact, it may be likely that mergers act as 
triggers for significant accretion \citep{AN02,DiMSH05, SDiMH04}.
Therefore, growth via mergers and accretion may be closely related.  
A detection of the gravitational wave background due to SBH 
mergers will directly address the frequency of SBH mergers as well 
as the BH masses involved in the mergers and hence 
constrain the relative importance of these events in 
galaxy formation. 

The key ingredients needed to calculate the consequences of 
hierarchical merging of seed BHs for LISA are 
the merger rate of BHs as a function of redshift and 
the distribution of BH masses involved in any 
particular merger event.  \citet{WL03}  calculated 
the merger rate using the extended Press-Schechter 
(EPS) approximation \citep{bond_etal91,lacey_cole93} 
convolved with the empirically-derived relationship 
between $\mbh$ and the velocity dispersion of the host halo.  
In a similar approach, \citet{RW05} 
derived a prescription for the black hole occupation distribution 
and showed that ringdown events will be observed in future 
gravitational wave detectors. \citet{MHN01} demonstrated how 
sensitive gravitational wave detection is to the merger rate of 
black holes. 
Another approach was taken in the studies of 
\citet{SETAL04a, SETAL04b} 
and \citet{islam_etal04a,islam_etal04b,islam_etal04c}, 
who populated halos with BHs according to a 
model like that of MR01 and followed 
the merger histories of these using the EPS formalism.  
\citet{EETAL04} adopted a different method, incorporating 
BH formation and evolution into a more elaborate, 
semi-analytic model of galaxy formation \citep{ENG03}.  
Another method for estimating merger rates was 
proposed by \citet{H94}, who developed phenomenological 
models for mergers based on the number counts of quasars 
and spheroids.  \citet{MSE04} showed that different BH assembly 
scenarios could yield different gravitational wave 
signatures with calculations based on the runaway 
SBH formation scenario of \citet{EETAL01}.

We present estimates of the merger rates and 
concomitant gravitational wave spectrum in a manner that is 
qualitatively similar to, but quantitatively different from the 
approach followed by \citet{SETAL04a} and  \citet{SETAL04b}.  We study 
the consequences of the KBD model of SBH formation 
which is markedly different from the MR01-type models.  
As we discuss in detail below, a basic prediction of the 
KBD model is a characteristic seed BH mass of order 
$\mbh \sim 10^5 \hMsun$, from which larger BHs 
must be built by mergers and/or accretion 
(as opposed to $\mbh \sim 10^2 \hMsun$ 
in the MR01-motivated models).  The strain induced 
during a merger event increases with the masses of the 
merging BHs while the characteristic frequencies decrease 
with BH mass (precise dependencies depend on the 
phase of the merger \citep[see for example ][]{FH98}).  
It seems reasonable then, that these scenarios result in 
distinct gravity wave signals.  We make specific 
estimates of BH merger rates by populating halos 
at high redshift with BHs according 
to the KBD prescription and we model 
subsequent dynamics using an approximate, 
semi-analytic formulation presented by 
\citet{ZB03} and expanded upon and tested against a 
suite of $N$-body simulations by \citet{ZETAL04}.  
We show that a robust prediction of the KBD scenario 
is a large gravity wave background.  The typical 
energy density per logarithmic frequency interval, 
in units of the critical density of the 
universe $\rhocrit$, has a peak value of 
$\Omegagw(f) \sim 10^{-7}$, which is achieved at 
frequencies $f \sim 10^{-4}-10^{-2} \si$, in an ideal 
range for the LISA interferometer.

The outline of this manuscript is as follows. In 
\S~\ref{section:kbdmodel}, we provide a brief outline 
of the KBD model of seed BH formation.  
In \S~\ref{section:merger_rate}, 
we step through our algorithm for estimating 
BH merger rates and present several illustrative, 
intermediate results that are useful for understanding 
the typical gravity wave signals.  
In \S~\ref{section:gravitationalwaves}, we briefly 
review the emission of gravitational waves by black 
hole mergers and present our calculation of the 
cosmological gravity wave background due to mergers of 
KBD seed BHs.  We present our results in 
\S~\ref{section:results}.  We summarize our results and 
present our conclusions in \S~\ref{section:conclusions}.  

We perform all of our calculations in the context of a 
standard, flat $\Lambda$CDM cosmology with 
$\Omega_{\rm M}=0.3$, $\Omega_{\rm \Lambda}=0.7$, 
$h=0.7$ and a power spectrum normalization of 
$\sigma_8=0.9$ \citep{tegmark_etal04}.  In what follows, 
we use the specifications of the proposed LISA instrument 
as an example of a gravity wave detector operating in the 
low-frequency regime.

\section{THE FORMATION OF SUPERMASSIVE BLACK HOLE SEEDS}
\label{section:kbdmodel}

In the scenario for SBH assembly proposed by KBD, 
BH seeds form from the lowest-angular momentum 
baryonic material in proto-galactic disks.  
Unlike the model of \citet{EL95}, 
where seed BHs are formed in the 
centers of dark matter halos with low {\em net} angular 
momentum, KBD proposed that seed BHs form from 
the lowest-angular momentum material in all halos 
that are able to assemble self-gravitating disks. 
Here, we give a brief overview of this mechanism 
for the generation of seed BHs and 
refer the reader to \citet{KBD04} for more details.

In the standard scenario of galaxy formation, galaxies 
form in halos that are massive enough to trap and cool baryons 
efficiently \citep{white_rees78}.  
The host dark matter halos acquire their angular 
momenta through tidal torques induced by neighboring 
quadrupole fluctuations in the density field.  
Baryons are assumed to have an angular momentum 
distribution that mirrors that of dark matter at very high redshift.  
Both \citet{VDBETAL02} and \citet{chen_etal03} 
tested this assumption using hydrodynamic cosmological 
simulations, and found it to be approximately correct.  
We still lack a detailed understanding of the 
evolution of the angular momentum distribution of 
baryons, yet the assumption that gas follows dark matter 
at high redshift is reasonable.

Unlike the dark matter, baryons are collisional.  Once 
trapped in dark matter potential wells, baryons cool by 
radiating their kinetic energy and fall further down the 
potential \citep[e.g. see][for a summary]{tegmark_etal97}.  
At high temperatures ($T \gtsim eV$), cooling 
proceeds via Bremsstrahlung radiation and collisional 
excitations.  Further cooling requires the presence of 
molecular hydrogen. Halos that cool efficiently 
are sufficiently massive that the production rate 
of molecular hydrogen is large enough to induce cooling 
on a timescale much shorter than a Hubble time. 
Once baryons begin to cool, they agglomerate at 
the halo center, and are expected to form a 
high-density, proto-galactic disk.  
The surface density of the disk is set by the 
angular momentum distribution of baryons. 
We assume an initial angular momentum distribution 
proposed by \citet{B01} (B01 hereafter).  
%
\begin{figure}[t]
\begin{center}
\includegraphics[height=7.2cm]{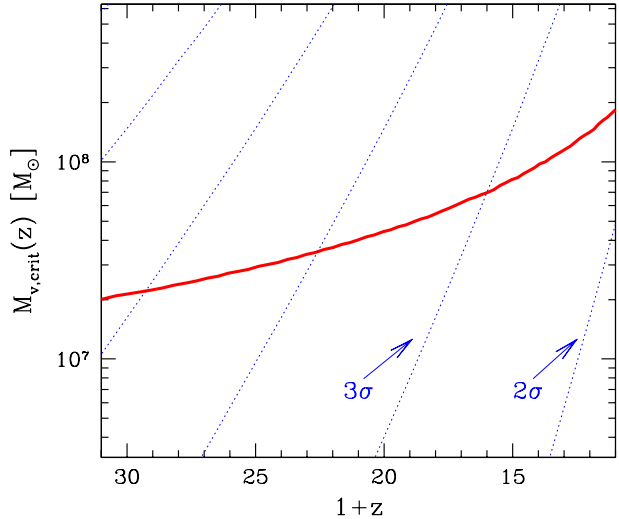}
\caption{
The critical mass for seed black hole formation as a 
function of redshift (solid line) in the KBD model. The dotted lines
correspond to 2$\sigma$-7$\sigma$ fluctuations in the 
smoothed density field indicating that the critical masses of 
concern are rare systems.
}
\label{fig:KBD_M_crit}
\end{center}
\end{figure}
In halos that efficiently cool their gas, 
a proto-galactic disk, with a surface density 
profile set by the B01 distribution, 
forms within roughly a dynamical timescale.  
Early on, this disk is pressure-supported. 
As more mass is added to the disk, the presence of local 
gravitational instabilities starts to overcome the pressure 
support.  In the inner parts of the disk, where such processes 
likely occur due to the higher densities, 
gravitational instabilities manifest themselves as an 
effective viscosity \citep{LP87}.  This viscosity 
leads to angular momentum transfer from the inner disk outward.  
Some material in the inner disk loses angular momentum resulting 
in the flow of mass to the center.  Eventually, instabilities 
in the disk become fully dynamical and the first generation of 
stars forms.  These first stars are thought to be massive, with 
lifetimes of order $\tfs \sim 1-10$Myr \citep{BHW01, HW02, SS02}. 
The presence of these first stars and their subsequent 
supernovae explosions likely result in significant heating of the disk. 
As the temperature of the disk increases, angular momentum transfer 
ceases to be effective and mass inflow stops.  Therefore, once 
gravitational instabilities become effective, the accumulation 
of mass at the center of the halo proceeds for a time $\sim \tfs$ 
and then stops.  

The above scenario assumes that the disk forms and 
evolves uninterrupted over many dynamical times.  However, 
a major merger is likely to disrupt any disk and halt 
this process prematurely.  Thus this process proceeds 
in halos that are both large enough to cool their gas 
{\em and do not experience major mergers}.  In practice, 
the latter criterion proves most stringent.  We estimate 
the major merger timescale as the average time for a halo 
to double its mass according to the EPS approximation, 
then we require this timescale to be longer than the 
time required to go through the above process.  
This sets the effective lower limit on the mass 
of halos that may form seed BHs.  We depict 
this {\em critical mass} $\Mcrit(z)$, for seed BH 
formation as a function of redshift as the {\em solid} 
line in Figure~\ref{fig:KBD_M_crit}.
In addition, we show contours in the Mass-redshift 
plane for $2-6\sigma$ fluctuations in the 
smoothed density field (using a standard top-hat window) 
to illustrate that this process proceeds in rare halos. 

By the time the viscous transport process terminates, 
a baryonic mass of order $\sim 10^5 \hMsun$ is 
transfered to the center of the super-critical halos. 
Such an object may be briefly pressure-supported, but 
it inevitably collapses to form a BH due 
to the post-Newtonian instability \citep{ST83}.  
In this regard, the KBD model distinguishes 
itself from models based on the proposal of \citet{MR01}.  
The KBD model predicts a characteristic minimum mass for the 
first, seed BHs of order $\mbh \sim 10^5 \hMsun$. 
In detail, the distribution of angular momenta exhibited 
by simulated cold dark matter halos manifests itself 
as a distribution in the masses of seed BHs formed via 
the KBD mechanism. Given a host halo mass 
$\Mvir \ge \Mcrit$ at $z$, 
the probability that a seed BH of mass $\mbh$ is 
formed within this halo is given by 
\begin{eqnarray}
\label{eq:Pmbh}
P( \mbh ) \, d \mbh &=& \frac{1}{\sqrt{2 \pi}} \, \frac{1}{\sigma_{{\mbh}}} 
\nonumber \\
& \times &  {\rm exp} \left[ - \frac{ \ln^2(\mbh / \mbh^0)}{
2 \sigma_{\mbh}^2} \right] \, d \mbh
\end{eqnarray}
where $\mbh^0$ is 
\begin{eqnarray}
\mbh^0 &=& 3.8 \times 10^5 \Msun \, \left( \frac{\Mvir}{10^8 \Msun} \right) 
\left( \frac{\fcb}{0.03} \right)^{3/2} 
\nonumber \\
& \times &
\left( \frac{1+z}{18} \right)^{3/2} 
\left( \frac{ \tfs }{ 10 {\rm Myr} } \right)
\end{eqnarray}
and $ \sigma_{\mbh} = 0.9$ \citep{KBD04}.  
In Equation~\ref{eq:Pmbh}, $\fcb$ is the 
fraction of cold baryons that form the 
self-gravitating disk in time $\tfs$. This is some fraction 
of the total possible baryonic material that can cool $f_0$, 
which we take to be roughly $\sim 20 \%$ of the universal 
baryon fraction in a $\Lambda$CDM cosmology.

The epoch  of cosmological  reionization places a  lower limit  on the
redshift  at   which  this  process  can  operate.    Heating  of  the
intergalactic medium (IGM)  prevents molecular hydrogen production and
essentially cuts  off the  accretion and cooling  of baryons  in small
systems.  Thus,  any significant IGM heating, such  as that associated
with   reionization,  effectively   stops  KBD   BH   formation.   For
simplicity, we  assume that  seed BH formation  stops abruptly  and we
refer to the redshift at which the KBD scenario effectively terminates
as $\zre$\footnote{In the context of  the KBD model, the inferred mass
density of black holes at $\zre$  allows for growth via accretion by a
factor of  $\sim 10$ in  mass, consistent with the  growth experienced
during an AGN phase of order  a Salpeter time in duration.  This is in
contrast to the model of \citet{VHM03} where in order to reproduce the
observed mass  density of  black holes, the  small mass seeds  of that
model  are expected  to  experience significant  accretion after  {\it
each} major merger experienced by their host halo.}.

In summary, the KBD model contains three key input parameters: 
the total fraction of mass that can cool via molecular hydrogen, $f_0$; 
the timescale for the evolution of the first stars, $\tfs$; 
and the redshift at which reionization heats the IGM to a 
degree that this process is no longer effective, $\zre$.  
Varying these parameters within a reasonable range, 
the predictions of the KBD model are most sensitive to $\zre$.  
This can be seen from Figure~\ref{fig:KBD_M_crit}.  $\zre$ sets 
the final redshift at which this process is effective 
and varying this redshift changes the rarity of the 
halos in which this process can take place.  In what 
follows, we adopt typical, fiducial values of 
$f_0 = 0.2$ and $\tfs = 10$ Myr and present results for 
$\zre = 8$, $\zre = 12$ and $\zre = 16$.  
Lower values of $\zre$ lead to more BHs and 
larger gravity wave signals, while it is interesting 
to explore high values of $\zre$ because 
they represent a minimum expected signal in this 
type of scenario and a high $\zre$ is supported by 
the cosmic microwave background anisotropy 
temperature-polarization cross correlation 
\citep{spergel_etal03,page_etal03}.

\section{ESTIMATING BLACK HOLE MERGER RATES}
\label{section:merger_rate}

In order to compute the gravitational wave signal due to 
BH mergers, we must compute the merger rate of black 
holes as a function of redshift and BH mass.  We do this 
in three steps.  First, we use the results of the KBD model, 
described in \S~\ref{section:kbdmodel}, to establish a population of 
seed BHs in host halos at high redshift.  Second, 
we estimate the rate at which BHs pairs are brought 
into the central region of a common halo center.  
\begin{figure}[t]
\begin{center}
\includegraphics[height=15.0cm]{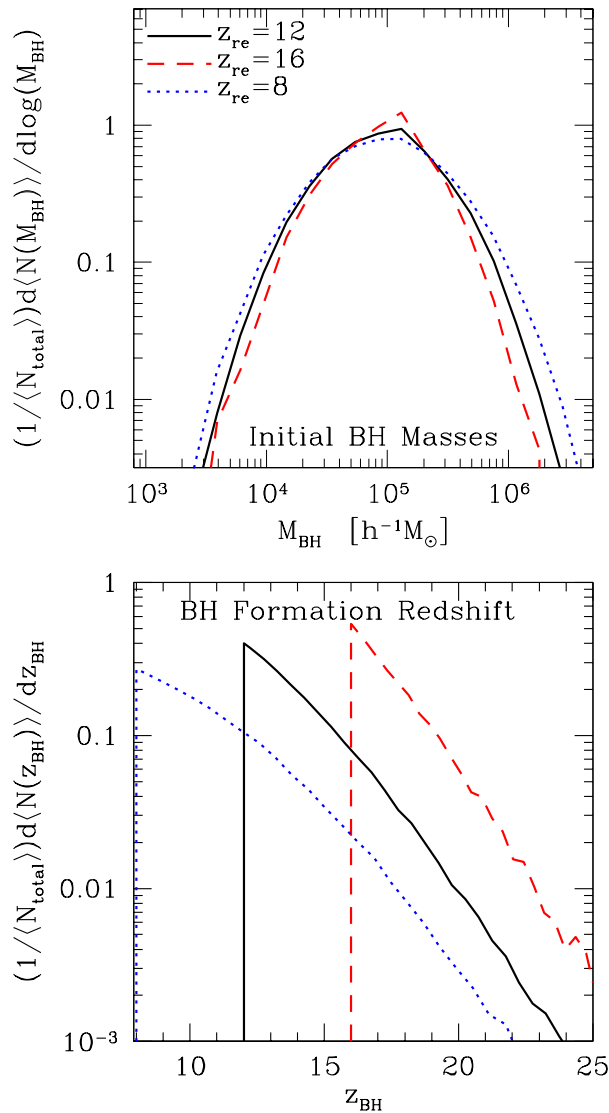}
\caption{
{\it Top}: The initial seed BH mass spectrum.  At 
the top of the panel we show the mean total number of black 
hole seeds per merger history realization in each case.  
These numbers provide the normalization for the curves.  
{\it Bottom}: The distribution of seed BH formation 
redshifts.  In both panels, {\it solid} lines correspond to 
our fiducial case of $\zre=12$ while {\it dashed} lines correspond 
to $\zre=16$ and {\it dotted} lines correspond to $\zre=8$.  
The sharp cutoffs represent the end of BH seed formation at 
$\zre$.  The results in both panels are from $200$ realizations 
of the formation history of a halo of mass $\Mvir = 10^{12.1} \hMsun$, 
which is comparable to the mass of the Milky Way halo.
}
\label{fig:bhi}
\end{center}
\end{figure}
To compute the rate of formation of these {\em potential pairs}, 
we use a semi-analytic model for halo mergers 
and the dynamical evolution of subhalos within larger host halos.  
We refer to this as the {\em halo evolution} stage.  These first 
two ingredients are based on the semi-analytic model of halo 
substructure evolution presented in \citet{ZB03} and \citet{ZETAL04}.  
Once potential pairs are formed within a halo, 
we estimate the timescale over which the 
BHs move from a separation of order a large fraction of a 
$\kpc$ to a separation close enough to emit 
gravitational radiation.  
We refer to this stage as {\em migration}.

\subsection{Populating Halos with Black Hole Seeds}
\label{sub:pophalos}

The first step is to establish a primordial population 
of seed BHs predicted by the KBD model.  Consider a halo 
of virial mass $\Mvir$ at $z=0$.  We use the extended 
Press-Schechter formalism \citep{bond_etal91, lacey_cole93} 
to compute merger histories, which consists of a list of all of the 
smaller halos that merged together to form this final halo and 
the redshifts at which these mergers occurred.  
This is a Monte Carlo procedure and by repeating the procedure 
many times for the same host mass, we sample the variety of 
different accretion histories that lead to a halo of 
fixed mass at $z=0$.  We refer to each sample merger 
history that we construct as a merger history {\em realization}.
In an individual realization, 
each merger represents a branch in the {\em merger tree}.  
At each branch, each merging halo has its own merger history 
that we can follow and so on.  The resulting merger tree 
structure describes the ways in which a halo 
of mass $\Mvir$ today breaks up into {\em halos} of various masses 
as a function of redshift.  We use the particular implementation of 
the EPS formalism advocated in \citet{somerville_kolatt99}, 
to which we refer the reader for further details.  
\begin{figure*}[t]
\begin{center}
\includegraphics[height=7.cm]{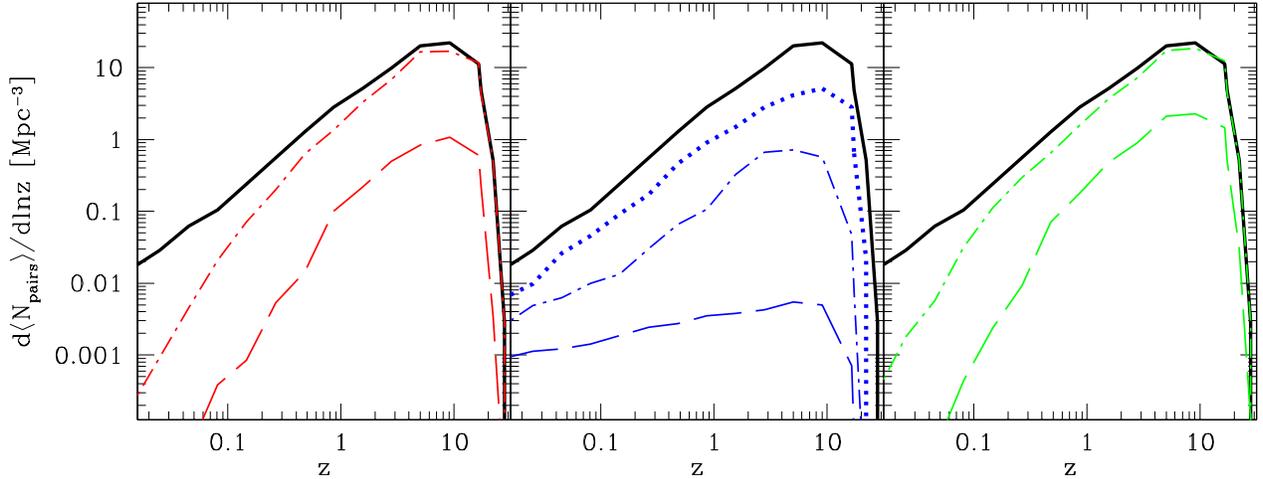}
\caption{
The mean comoving number density of potential BH pairs as a function of 
redshift for the fiducial case of $\zre=12$. 
In all three panels, the {\it thick, solid} curve represents 
the total number density of potential BH pairs. {\it Left}: 
The {\it dot-dash} line represents the total number density 
of potential 
BH pairs for which the total mass in the binary is 
$M_1+M_2 \le 10^6 \Msun$, while the {\it dashed} line depicts all 
pairs with $M_1+M_2 \le 10^5 \Msun$. 
{\it Middle}: The {\it dotted} line represents the number density 
of pairs where the ratio of black hole masses is $M_2/M_1 \le 10^{-1}$, 
the {\it dot-dashed} line shows the pairs with $M_2/M_1 \le 10^{-2}$, 
while the {\it dashed} line corresponds to pairs with 
$M_2/M_1 \le 10^{-3}$. {\it Right}: Here, the {\it dot-dashed} line 
represents pairs where the most massive black hole mass is 
$M_1 \le 10^6 \Msun$, while the {\it dashed} line depicts pairs where 
$M_1 \le 10^5 \Msun$. 
}
\label{fig:N_z}
\end{center}
\end{figure*}

We follow each branch to higher redshift and smaller mass until 
one of two things happens.  If the branch is above $\Mcrit(z)$ at 
some redshift $z > \zre$, then we find the earliest time in 
its mass accretion history that the the halo mass was 
greater than $\Mcrit(z)$.  At this time we assign this the 
redshift of BH formation, $\zbh$.  
We choose the earliest time because we assume that subsequent 
mergers between super-critical halos do not restart the process 
described in \S~\ref{section:kbdmodel}.  This is because once 
a halo goes through the mechanism of BH formation described 
in \S~\ref{section:kbdmodel}, the presence of metals left over from 
the first generation of stars in the disk alters the properties of 
the disk, making it significantly more susceptible to fragmentation.  
We then assign the halo a BH mass 
according to the probability distribution in 
Equation \ref{eq:Pmbh}.  Alternatively, if there is no 
point on the branch where $M > \Mcrit(z)$ at $z > \zre$, 
we eliminate this branch from further consideration.  
In this case, the halo does not meet the conditions 
necessary for the formation of a SBH seed and 
can not host a BH.  

As an example of the product of this procedure, 
in Figure~\ref{fig:bhi} we show 
initial spectra of seed BH masses and 
formation times.   
In this specific demonstration, 
we calculate these quantities by 
averaging the results of $200$ merger histories of halos 
of final mass $\Mvir = 10^{12.1} \hMsun$, comparable 
to the halo of the Milky Way \citep{klypin_etal04}, 
with our fiducial value of $\zre=12$, 
as well as low and high reionization 
redshift alternatives $\zre=8$ and $\zre=16$. 
The mean total number of seed black holes is  
$\langle {\rm N_{total}} \rangle = 589$ for the fiducial 
case of $\zre=12$, while for $\zre=8$ and $\zre=16$ this number
is $\langle {\rm N_{total}} \rangle = 1601$ and 
$\langle {\rm N_{total}} \rangle = 122$ respectively.
The spectra in Figure~\ref{fig:bhi} 
are normalized to the mean number of seed BHs. 
First, notice that the formation redshifts are peaked at 
$\zbh = \zre$ in all cases.  
Figure~\ref{fig:KBD_M_crit} shows halos 
that are suitable hosts for BH seeds become increasingly rare as 
redshift increases with the result that most BHs form at 
$z \sim \zre$.  
Additionally, the initial BH mass spectrum is 
a broad distribution. In our fiducial case, it is roughly 
log-normal with $\langle \log (\mbh/\hMsun)\rangle \approx 5.4$ 
and a large dispersion, $\sigma(\log_{10} (\mbh) ) \approx 0.45$.  
This mass distribution is broader than that of Eq.~\ref{eq:Pmbh}, 
due to the variety of different halos and redshifts at which 
these seed BHs may form.  The distribution 
gets systematically narrower with increasing $\zre$ 
because the halos suitable for seed BH formation become 
increasingly rare so that BH formation occurs in correspondingly 
narrower ranges of host halo mass and formation redshift.

\subsection{Halo Evolution}
\label{sub:haloevol}

With initial BH populations in place, the next step is 
to compute the evolution of the BHs and their host halos, 
as they merge.  We do this using a semi-analytic model 
for the merging and evolution of substructure in a larger main halo 
described in detail by \citet{ZETAL04}.  In this section, 
we outline this model.

First, merger trees are constructed as described in \S~\ref{sub:pophalos}.  
Each halo of mass $\Mvir$ today forms from a sequence of mergers with 
halos of smaller mass.  We refer to the larger of the pair as the 
{\em main halo} and the smaller as a {\em subhalo}.  BH 
mergers are always preceded by these halo mergers as it is the 
halos that bear the infalling BHs. 
BHs are brought near the center of the main halo, 
where they presumably merge, primarily 
due to dynamical friction acting to remove orbital energy 
from the orbiting subhalo.  Roughly speaking, the deceleration due to 
dynamical friction is proportional to the bound mass of the subhalo, 
so the loss of mass due to the tides raised by the potential of the 
main halo must also be modeled in order to make an 
estimate of the timescale on which subhalos bring 
BHs to the central regions of the main halo.  

We track the evolution of subhalos as they evolve in the growing, 
main system in the following way.  
At each merger event, we first assign initial orbital 
energies and angular momenta to infalling subhalos according to 
the probability distributions culled from cosmological 
$N$-body simulations \citep{ZETAL04}.  
We then integrate the orbit of the 
subhalo in the main halo from the time of accretion 
to the present day.  
We model tidal mass loss using a modified 
tidal approximation and dynamical friction using 
a modified form of the Chandrasekhar formula 
\citep{chandrasekhar43}, as described in detail 
in \citet{ZETAL04}.  For simplicity, we model 
all halos by the density profile of 
\citet{nfw97} (NFW).  \citet{ZETAL04} showed that this 
model is in excellent agreement with the results of direct 
simulation over several orders of magnitude in halo mass 
and over a wide range of redshifts.  The only modification 
that we make to the prescriptions in \citet{ZETAL04} is that 
halos may have an additional point mass representing 
the BHs at their centers.  In practice, 
the presence of the BHs make little difference 
in the large-scale dynamics because the BH 
masses are $\ltsim 10^{-3} M_{\mathrm{subhalo}}$.  
\begin{figure}[t]
\begin{center}
\includegraphics[height=13cm]{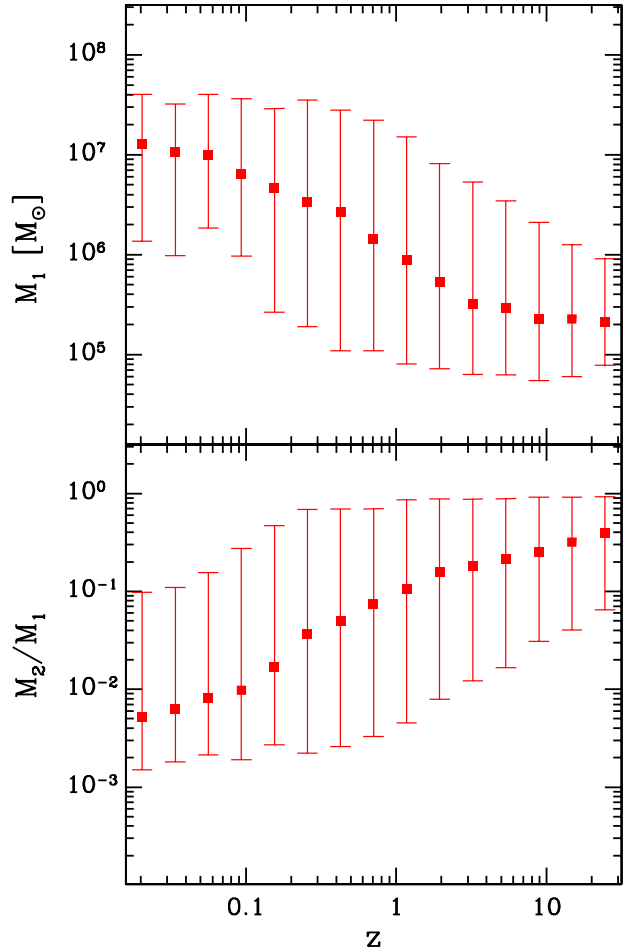}
\caption{
{\it Top}: The median mass of the heaviest BH in a pair 
as a function of redshift. The error bars represent the 
90 percentile range of black hole masses.  
{\it Bottom}: The ratio of the lighter to the heavier black 
hole mass $M_2/M_1$, in a pair as a 
function of redshift. Errorbars are as in the top panel.
}
\label{fig:M1_MR_z}
\end{center}
\end{figure}
Subhalos that approach the central 
$\sim \mathrm{few }\  \kpc$ of the main halo 
lose most of their mass due to tides, yet their 
BHs continue to orbit the host.  Once the apocenter 
of an orbiting system becomes very small, we cannot continue 
to track its evolution for two reasons.  First, the dynamical 
timescales in the central halo are quite short so that 
very small timesteps are needed to integrate the orbit and the 
calculation becomes unmanageable.  Second, our simple model 
of subhalo evolution is not applicable near the centers 
of halos where interactions with the central galaxies 
often dominate interactions with halo potentials.  
We follow the approach of \citet{islam_etal04a} and 
cease tracking the orbits of all subhalos and BHs that 
reach an apocentric distance from the center of the 
main halo that is less than 
$\rpp = \min( 0.01\Rvir, 1 \kpc )$, 
where $\Rvir$ is the virial radius of the  main halo.  
We refer to the resultant pairs that are separated by 
$\sim \rpp$ as BH {\em potential pairs}.  
In this way, we decouple the large-scale subhalo dynamics 
that bring the BHs from a separation of $\sim \Rvir$ 
to a distance $\ltsim 0.01 \Rvir$, from the 
subsequent processes that determine the rate at 
which potential pairs evolve into binaries and merge.  
We discuss this final stage of BH 
{\em migration} in the following subsection.

For the purposes of our modeling here, 
we compute $25$ merger tree and subhalo evolution 
realizations at $15$ values of the 
$z=0$ host mass $\Mvir$, 
starting at $\Mvir = 10^{11}\hMsun$ 
and increasing in steps of 
$\Delta [\log (\Mvir)] = 0.25$.  
We find that extending the computation 
to smaller masses does not influence our results 
because smaller halos generally host few BHs, while 
extending our computation to higher masses does not 
affect our results because of the relative rarity of such 
large halos.  

We show an example of the intermediate results after 
this second stage of modeling in Figure~\ref{fig:N_z}. 
In this Figure, we plot the mean comoving number density 
of black hole pairs as a function of redshift. At high 
redshift and up to the redshift of reionization there 
is a linear increase of BH pairs as new seed black holes 
are being formed (see discussion in Section~\ref{section:kbdmodel}). 
The increase in potential pairs comes from the fact that 
halos that are able to host seed black holes become 
increasingly more abundant with decreasing  redshift. 
At reionization seed BH formation ceases to operate and 
the number of BH pairs is governed solely by the interplay 
of halo merger rates and dynamics.  The former depend 
on the underlying cosmological model, while 
the latter depend on the details of 
dynamical friction, tidal heating, and mass loss. 

Figure~\ref{fig:M1_MR_z} depicts the masses of the black 
holes that form the potential BH pairs. As expected, 
at early times, prior to significant merger activity, 
the masses of BHs in the centers of all halos are of the same 
order as the seed BH masses in Eq.~(\ref{eq:Pmbh}).  
As major mergers bring more BHs to the center, 
the masses of BHs in large halos increases.  
Mergers of small, satellite black 
holes with larger, central BHs that are themselves 
the products of mergers become more common and 
drive the ratio $M_{2}/M_{1}$ down with decreasing redshift.  
The relatively greater importance of these mergers with 
remnants of previous mergers at low redshifts 
can also be gleaned from Fig.~\ref{fig:N_z}.
Again, we emphasize that 
the number of BHs and BH pairs is sensitive to $\zre$, 
yet the masses are not because $\Mcrit$, 
and consequently the typical BH mass $\mbh^0$ 
varies very slowly with redshift 
[see Fig.~\ref{fig:KBD_M_crit} and Eq.~(\ref{eq:Pmbh})].

\subsection{Black Hole Migration}
\label{sub:migration}

As we showed in the previous subsection, we track BHs until the 
distance between the parent BH and the merging BH is 
$\rpp = \min( 0.01 \Rvir, 1 \kpc )$.  This distance is 
much larger than the distance where gravitational radiation  
becomes an important channel for energy loss from the pairs.  
Somehow, the infalling satellite BHs in the potential pairs 
must {\em migrate} to the center.  
This can be achieved either through stellar interactions 
\citep{BBR80,Q96,MM01,Y02}, or various gas dynamical effects 
\citep{GR00, AN02, kazantzidis_etal04}.  
\citet{SETAL04a, SETAL04b} calculate the timescale for the pair to 
reach the gravitational radiation regime due to stellar interactions, 
following the analysis of \citet{Q96}.  
In order to implement such a scenario, 
they make the mapping between velocity dispersion and 
halo circular velocity based on an empirical 
relationship \citep{F02}.  The subsequent hardening 
of the binary proceeds through the capture and ejection of  
nearby stars.  Such estimates require numerous uncertain 
assumptions.  Instead, we choose to leave the migration 
timescale $\tmig$ as a parameter, 
and investigate the consequences of various migration 
timescales on our final results.

We can make very rough estimates of the range of values 
that $\tmig$ may
plausibly take.  
The fastest timescale whithin which a black hole 
would be
able to migrate to the center of the halo is the 
dynamical timescale.  If
we assume black holes are at a distance $\rpp$ 
from the center, then
typical values of the dynamical timescale are of 
order $\sim$ Myr for the
halos that host black holes in our model.  
Black holes may sink to the center by exchanging 
their orbital 
energy with nearby gas, stars and dark matter. 
If we follow the
method of \citet{SETAL04a, SETAL04b}, we find 
that the migration timescale
ranges from $\tmig \sim 1$~Myr to a few Gyr, 
depending upon the initial
configuration of the pair ($\rpp$), the properties 
of the host halo
($\Mvir$), and the central stellar population.  
The latter can be obtained
given the mass of the host halo and the 
empirically-derived relationship
between stellar velocity dispersion and halo 
circular velocity
\citep{F02}.  We assume $\tmig$ takes values 
that range from $\tmig =
1$~Myr to $\tmig = 1$~Gyr, to cover a range of 
reasonable expectations.
Migration timescales in excess of a $\tmig \sim 1$~Gyr 
will most likely be
altered by other processes because the major merger
 timescale is of the
same order. In this case, the pair may get disrupted, 
and/or following a
potential 3-body interaction, the characteristics 
of the pair may change
(for a comprehensive review of the evolution of 
BH binaries see 
\citet{MM04}).

The net effect of changing the value 
of $\tmig$ is easy to understand in general. 
If we choose the lower bound, $\tmig \sim$~Myr, 
almost all pairs will reach the gravitational 
emission regime by $z=0$. On the other hand, if 
$\tmig \sim$~Gyr, pairs that are formed 
at redshifts smaller than $z \ltsim 0.1$ will not 
reach the gravitational emission regime by today. 
However, as can be seen in Figure~\ref{fig:N_z}, 
the fraction of mergers occurring at $z \ltsim 1$ is 
rather small so the effect on the cumulative 
gravitational wave background is small.  
Thus, the choice of $\tmig$ is {\em not} crucial 
to our results. 

\section{GRAVITATIONAL WAVE EMISSION FROM BLACK HOLE MERGERS}
\label{section:gravitationalwaves}

\subsection{Gravity Waves From Individual Mergers}
\label{sub:gw_binary}

In this section, we briefly review the emission of 
gravitational waves from coalescing BH pairs.  
Most of our review follows from the comprehensive 
discussions in \citet{misner_etal73}, 
\citet{thorne80}, \citet{thorne87}, 
\citet{FH98, flanagan_hughes98b},and 
\citet{cornish_larson01}.  We refer the reader 
to these resources for details.  

As in electromagnetism, the gravitational 
radiation field can be decomposed into multipoles and the 
energy output is typically dominated by the lowest 
non-vanishing multipole.  In the case of electromagnetic 
radiation, this is usually the dipole, whereas gravity is a 
tensor field so the lowest non-vanishing term is typically 
the quadrupole.  The power output in quadrupolar radiation from a 
slowly-moving, weak-field gravity wave source 
is then similar to the familiar far-field relation for 
electromagnetic quadrupole radiation 
\citep{jackson75} and is given by 
\citep[e.g.][]{misner_etal73} 
\begin{equation}
\label{eq:dedt}
\frac{dE}{dt} = \frac{c^5}{G}
\langle \frac{ d^3 I_{jk} }{ dt^3 } \frac{d^3 I^{jk} }{dt^3} \rangle,  
\end{equation}
where 
\begin{equation}
\label{eq:inertia}
I_{jk} = \int  \rho 
\left( x^jx^k - \frac{1}{3} r^2 \delta_{jk} \right) \, \dd^3x
\end{equation}
is the moment of inertia of the mass distribution 
and overdots denote time derivatives. 

Rather than flux, gravity wave detectors measure 
the wave amplitude through the fractional 
change in the physical separation of 
two masses, expressed as the strain $h(t)$.  
The metric perturbation due to the wave $h_{ \mu \nu }(t)$, 
contains two polarizations, which in the 
transverse-traceless gauge, are conventionally labeled as 
$h_+(t)$ and $h_{\times}(t)$.
If we define the Fourier transform of $h_i(t)$, 
with $i=+,\times$, as 
$\tilde{h}_i(f) = \int_{-\infty}^{+\infty} h_i(t) \, e^{2 \pi i f t } \, dt$, 
then the one-sided spectral density of gravity waves 
$S_h(f)$ is  defined as 
$2 \sum_i \langle \tilde{h}_i^*(f) \tilde{h}_i(f') \rangle = 
\delta( f - f') \, S_h(f)$. 
The dimensionless characteristic strain amplitude is 
defined by $h_c^2(f) = f S_h(f)$.  Averaging over 
polarizations of the wave and the orientations of the 
source relative to the detector, the characteristic 
strain amplitude from a merger at redshift $z$ is \cite{FH98}
\begin{eqnarray}
\label{eq:hoff}
h_c(f)  &=& \sqrt{ \frac{ 2 G }{ c^{3} } } 
\frac{ (1+z)}{\pi d_L(z)} 
\sqrt{ \frac{dE}{df_s} } \nonumber \\
&=& 3.56 \times 10^{-26} \,
\left( \frac{{\rm Mpc}}{d_L(z)} \right) 
\left( \frac{1+z}{1+10} \right) \,
\sqrt{ \frac{dE/df_s}{M_\odot {\rm m}^2 \si}},
\end{eqnarray}
where $d_L(z)$ is the luminosity distance to the merging 
BH pair,
$dE/df_s$ is the energy spectrum of the radiation at the 
source, and $f_s = f(1+z)$ is the frequency at the source.
\begin{figure}[t]
\begin{center}
\includegraphics[height=7.cm]{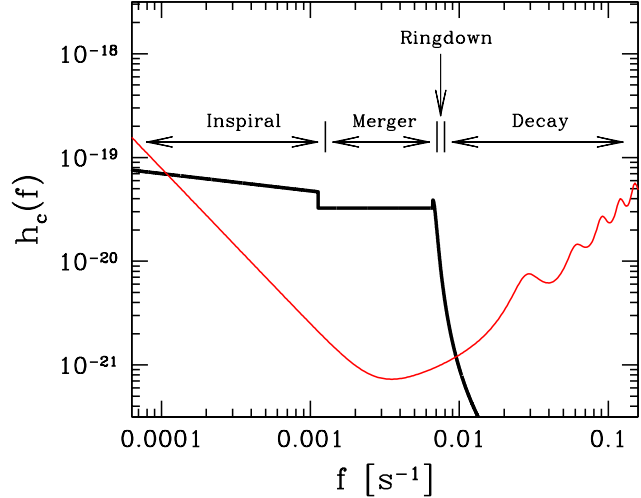}
\caption{
An example of the strain induced by a 
high-redshift black hole merger.  
The different merger phases in a pair 
of masses $M_1 = 1.5 \times 10^5 \Msun$ and 
$M_2 = 1.1 \times 10^5 \Msun$ that initiates 
gravitational wave emission at $z = 13$ 
({\em thick line}).    The 1$\sigma$ 
detection threshold for burst sources 
with LISA is shown as the {\em thin} curve 
\citep[e.g.][]{thorne87,thorne96,larson_etal01}.  
}
\label{fig:sample_h}
\end{center}
\end{figure}
There are four phases of gravity wave emission during the 
coalescence of a BH pair and we now discuss the energy 
spectra, $dE/df_s$, of gravity waves emitted during 
each phase in turn.  We refer to the masses of the two BHs as 
$M_1$ and $M_2$, with $M_1 > M_2$.  Following \citet{FH98}, 
we refer to the four phases of BH coalescence as {\em inspiral}, 
{\em merger}, {\em ringdown}, and {\em decay}. 
We show these phases schematically in Figure~\ref{fig:sample_h} 
for a single merger of BHs of mass $M_1 = 1.5\times 10^5 \Msun$ and 
$M_2 = 1.1 \times 10^5 \Msun$ at $z=13$.  These values are 
typical of the individual mergers in our calculations 
(see \S~\ref{section:merger_rate}).  The thick line in 
Fig.~\ref{fig:sample_h} represents the $5\sigma$ 
detection threshold for a $1$~year mission with 
LISA based on the sensitivity characteristics proposed for the 
LISA interferometer.  We generated this curve and all 
subsequent sensativity curves with the LISA 
{\em online sensitivity curve generator}
\footnote{This is available at URL 
{\tt http://www.srl.caltech.edu/\~{}shane/sensitivity/}}.
Signals just above threshold and with a well-understood waveform 
can be culled from the data stream using optimal filtering 
techniques \citep[e.g.][]{FH98}; 
however, it is already clear from Fig.~\ref{fig:sample_h} 
that if our model is correct, LISA should be able 
to study the predicted gravitational wave 
background at high signal-to-noise.  

First, during the {\it inspiral} phase, the two BHs are 
well-separated and lose energy by the self-interaction of 
the pair with the emitted waves.  This radiation-reaction drag 
causes the two BHs to inspiral slowly toward each 
other.  At this stage, the adiabatic, weak-field 
limit can be applied in a straightforward way, yielding
\begin{eqnarray}
\label{eq:dedf_inspiral}
\left( \frac{dE}{df_s} \right)_{{\rm Inspiral}} &=& 
\frac{1}{3} \Bigg( \frac{\pi^{2}G^{2}}{f_s}\Bigg)^{1/3} 
\frac{ M_1 M_2}{(M_1 + M_2)^{1/3}} \nonumber \\
&=& 1.86 \times 10^{13} \, M_\odot {\rm m}^2\si 
\left( \frac{f_s}{\si} \right)^{-1/3} \nonumber \\
&\times& \left[
\frac{M_1M_2}{(M_1 + M_2)^{1/3}} \right], 
\end{eqnarray}
where the BH masses are expressed in $\Msun$.  
The inspiral phase begins at a frequency (equivalently, 
a physical separation between BHs) where the 
emission of gravitational waves becomes the 
dominant mode of energy loss and ends at a 
frequency corresponding to the transition from 
radiation-reaction-driven inspiral of the two BHs 
to a free-falling plunge toward each other.  

We label the frequency at which inspiral is initiated 
as $\fmin$, and the frequency at which it ends as 
$\fmerger$. The value of $\fmerger$ is roughly set at 
$\fmerger \sim 4.1 \times 10^{-3} M_{\rm{total},6}^{-1} \, \si$ 
where $M_{{\rm{total}},6}$ is the total 
mass in the pair expressed in units of $10^6 M_\odot$. 
Even though the value of $\fmerger$ may vary up to several 
percent if the masses involved in the pair are unequal 
and/or if the BHs are rotating, we choose this 
conservative value in order to circumvent the 
need to generate numerical templates that would be 
required if the frequency were higher \citep[see ][]{FH98}.  
A low value of $\fmerger$ is conservative because it 
{\em minimizes} the energy emitted per frequency interval 
during the merger phase (see below), thus minimizing the strain 
induced during the merger.  The value of $\fmin$ is rather 
uncertain.  Previous studies assumed it to be a constant 
\citep{WL03}, or set it to the frequency at the 
so-called ``hardening'' radius of the system \citep{SETAL04a}. 
We choose to leave it a free parameter, and 
parameterize it in terms of the variable $\beta$, 
defined as $\beta \equiv \fmin / \fmerger$. 
We then examine the sensitivity of our results to the 
$\beta$ parameter. We assume that $\beta$ may take 
a range of reasonable values between $10^{-3} - 10^{-1}$.  
These values span the range that the masses of the 
BHs and host halo properties involved 
in our calculation imply, had we 
followed the same procedure as \citet{SETAL04a}.  

The time evolution of the emitted frequency of 
gravitational waves during inspiral is set by the 
evolution of the orbital frequency of the binary 
and is given by 
\begin{equation}
\label{eq:time_in}
t_{\rm inspiral}(f_1 \rightarrow f_2) \simeq 3.2\times 10^{-3} \, {\rm yr} 
\frac{ (M_1 + M_2 )^{1/3}}{M_1M_2} \left(
f_1^{-8/3} - f_2^{-8/3} \right)
\end{equation}
It is easy to predict the effect of changing $\beta$ on our results.  
Choosing $\fmin$ very close to the value of 
$\fmerger$ ensures that nearly all potential pairs complete the 
inspiral phase (unless of course they were accreted at very low 
redshift, see the discussion on $\tmig$ in \S~\ref{sub:migration}). 
In this case, the central BHs grow by the large number of 
merger events they experience.  On the other hand, if we choose 
$\beta$ very small (e.g. $\fmin \ll \fmerger$), 
then the time spent in the inspiral phase may, in some 
cases, approach the Hubble time and most pairs will not 
merge by $z=0$. In this case, the central BH does not 
grow significantly by mergers with incoming BHs. 

After the inspiral phase, a BH pair reaches the 
maximum frequency $\fmerger$.  The system 
transitions from an inspiral to a plunge toward 
each other.  This stage of evolution of a binary 
system is poorly-understood and, following \citet{FH98}, 
we refer to this as the {\em merger} phase. 
The merger is highly non-linear and numerical 
simulations of this process are not yet accessible.  
The merger phase is generally rapid 
(of order minutes, 
$ t_{\rm merg} \sim 1.47 \times 10^2 \, {\rm s} \, \, M_{\rm 6, total}$), 
and the system quickly evolves to the quasi-normal-ringing phase 
\citep[see the extensive discussion in][]{FH98}. 
The merger event occurs in the frequency interval 
between $\fmerger$ and $\fqnr$, where $\fqnr$ is the 
quasi-normal-ringing frequency set by the spin of the 
resulting BH.  The resulting BH may be rapidly rotating 
even if the individual BHs had very 
small or zero angular momentum \citep[however, see][]{HR03}. 
We assume a value for $\fqnr$ given by the analytic fit of \citet{E89} 
$\fqnr \approx c^3 [ 1 - 0.63 ( 1 - a )^{3/10}]/ 2 \pi G \,(M_1 + M_2)$, 
where $a$ is the dimensionless spin parameter of the resulting 
BH of mass $M_1 + M_2$. 
We assume that the spin parameter takes a value of $a = 0.95$, 
in order to minimize the resulting strain induced from the 
decay phase of gravitational wave emission (see below).  
The resulting value of $\fqnr$ is then 
$\fqnr = 2.41 \times 10^{-2} M_{\rm total,6}^{-1} \, \si$.  
We use an approximation for the energy spectrum 
from the merger phase presented by \citet{FH98}: 
\begin{eqnarray}
\label{eq:dedf_merger}
\left( \frac{dE}{df_s} \right)_{{\rm Merger}} &=& 
\frac{\epsilonm (M_1 + M_2) \, \xi(\mu, M_1, M_2 ) \, c^2 }
{ \fqnr - \fmerger }  \, \Theta_1 \, \Theta_2 \nonumber \\
&=& 2.7 \times 10^{15} \, M_\odot {\rm m}^2\si 
\left( \frac{ \epsilonm}{0.03} \right) \nonumber \\
& \times &
\frac{ (M_1 + M_2) \, \xi(\mu, M_1, M_2 ) }{ \fqnr - \fmerger } \, 
\Theta_1 \, \Theta_2 
\end{eqnarray}
where $\Theta_1 = \Theta (f-\fmerger)$ and 
$\Theta_2 =  \Theta (\fqnr - f) $ are step functions 
and the masses are in units of $\Msun$. 
The function $ \xi(\mu, M_1, M_2 ) $ 
is the reduction function which ensures correct 
fitting to numerical calculations for equal and 
non-equal mass BH pairs. It has the value 
$\xi(\mu, M_1, M_2 ) = [ 4 \mu / (M_1 + M_2 )]^2$, where 
$\mu$ is the reduced mass of the pair.  
The quantity $\epsilonm$ is the fraction of the 
total mass $M_1 + M_2$, 
released via gravitational radiation 
during the merger phase.  The efficiency at which 
energy is radiated depends on the mass 
of the BHs in the merging phase, 
their individual spins, and their orbital 
configuration just prior to the merger 
(generally taken to be the orbital 
parameters at $\fmerger$). 
\citet{FH98} estimate $\epsilonm$ based 
on angular momentum conservation.  
By assuming an idealized scenario where 
BH spins and orbital angular momenta are aligned, 
and assuming that only the quadrupole radiation 
carries away energy, they estimate a value of 
$\epsilonm \sim 0.1$. However, 
based on the approach of \citet{S79}, 
\citet{D79}  showed that the efficiency can be as 
low as $\epsilonm \gtsim 0.03$ for merging BHs with 
no spin.  To be conservative, we choose 
$\epsilonm = 0.03$ for the rest of our calculation.

The {\em ringdown}  phase corresponds to 
damped oscillations of the quasi-normal 
modes of the resultant BH at the 
quasi-normal-ringing frequency $\fqnr$. 
{\em Decay} is the subsequent emission of decaying 
gravitational waves as the merged remnant settles 
toward a Kerr BH \citep{E89}.  
The remnant settles toward a BH with 
mass that may approach the sum of the two masses 
$M \ltsim M_1 + M_2$.  The actual resulting mass depends upon 
on the details of the merger which, to date, are poorly understood.  
An estimate of the energy spectrum of gravitational waves 
during the ringdown and decay phases is \citep{FH98},
\begin{eqnarray}
\label{eq:dedf_qnr}
\left( \frac{dE}{df_s} \right)_{{\rm Ringdown}} &=& 
\frac{8 \, c^2}{32 \pi^3} \frac{ \epsilonr}{Q \, \fqnr} 
\left( \frac{f_s}{\tau} \right)^2 \nonumber \\
&\times&( M_1 + M_2 ) \,\xi(\mu,M_1,M_2) 
\, B(f_s, \fqnr) \nonumber \\
&\simeq& 9.1 \times 10^{11} \, M_\odot {\rm m}^2\si \, 
\left( \frac{\epsilonr}{0.01} \right) \nonumber \\
&\times& (M_1 + M_2)\,\xi(\mu,M_1,M_2) \\
&\times& \left( \frac{16}{Q} \right) 
\left( \frac{ f_s / \si}{ \tau / {\rm s }} \right)^2 
\frac{1}{\fqnr} B(f_s, \fqnr)
\end{eqnarray}
where masses are expressed in $\Msun$, 
$Q$ is the quality factor, 
and $\tau$ the damping time of the 
quasi-normal ringing mode 
(assumed to be the $l=m=2$ mode), 
and they are related by 
$Q = \pi \fqnr \tau \approx 2 ( 1 - a )^{-9/20}$ 
\citep{E89}.  The function $B(f_s, \fqnr)$ is defined as 
\begin{eqnarray}
\label{eq:Bqnr}
B(f_s, \fqnr) &=& \frac{1}{ 
[( f_s - \fqnr )^2 + ( 2 \pi \tau)^{-2} ]^2}   \nonumber \\ 
&+& \frac{ 1}{ [ ( f_s + \fqnr )^2 + ( 2 \pi \tau)^{-2} ]^2 } 
\end{eqnarray}

The amplitude of gravitational waves produced 
at ringdown depends weakly on the value of $a$.
However, the rate at which the induced strain falls 
off during the decay phase depends on $a$ through 
the damping time $\tau$. Specifically, 
the strain amplitude at $f > \fqnr$ is higher 
for smaller values of $\tau$.  The spin 
parameter generally depends on all of the details of 
BH formation and merger.  To be conservative and 
minimize the induced strain during merger, ringdown, 
and decay, we take a constant $a=0.95$. 
The quantity $\epsilonr$ is the fraction of energy 
released in gravitational waves during ringdown and 
decay.  The amplitude of the gravitational radiation 
during ringdown, and hence $\epsilonr$, are not well known. 
Numerical simulations give $\epsilonr \sim 0.03$ 
\citep{BS95}.  This value can also be obtained 
under simplifying assumptions about the origin of 
the waves during ringdown \citep{FH98}.  
For lack of any other estimates, and to remain conservative 
in our approach, we choose a constant $\epsilonr = 0.01$\footnote{During 
the preparation of this manuscript, \citet{SETAL05} reported the 
results of a new calculation based on mesh-refinement and 
dynamic singularity excision where they estimated the fraction of 
the total ADM mass of a merged system 
that is radiated during ringdown to be $\sim 0.1\%$.}.

\subsection{Black Hole Recoils from Asymmetric Gravity Wave Emission 
and Many Body Interactions}
\label{sub:kicks}

Using the methods described in 
\S~\ref{sub:pophalos} to \S~\ref{sub:gw_binary}, we have 
a complete description of the formation of BH seeds at high 
redshift, their evolution, and their merger histories 
notwithstanding a few physical parameters that contain 
our ignorance of various details.  There are two additional 
effects that serve to complicate the situation 
and it is necessary to examine the robustness 
of our results to these effects.  

One effect that has been neglected in all but one 
of the previous analyses \citep{SETAL04a} 
is that BH mergers result in a considerable
recoil of the center of mass of the system due to 
the asymmetric emission of gravitational waves 
during the final stages of coalescence \citep{fitchett83}.
\citet{favata_etal04} recently addressed this problem 
in detail and found that expected recoils could be 
\begin{equation}
\label{eq:vrecoil}
\vrecoil \sim [40-500] g(M_1/M_2)/g_{\mathrm{max}} \kms,
\end{equation}
where the dependence upon mass ratio is encapsulated in 
$g(x) = x^2(1-x)/(1+x)^5$ which reaches of maximum of 
$g_{\mathrm{max}} \simeq 0.018$ at 
$x_{\mathrm{max}} \simeq 0.382$.  
The uncertainty in the normalization reflects the 
uncertainty in the calculation of the recoil during 
the final ``plunging'' stage of BH coalescence and the 
range we quote corresponds to the {\em upper limit} 
and {\em lower limit} estimates of \citet{favata_etal04}.  
The function $g(x)/g_{\mathrm{max}}$ is broad about its 
maximum, such that $g(x)/g_{\mathrm{max}} \gtsim 0.1$ for 
$0.03 \ltsim x \ltsim 0.98$.  

The size of the recoil and the mass ratios for which this 
recoil is sizable are in a range such that they may have 
a significant effect on our results.  Figure~\ref{fig:N_z} 
shows that the majority of BH mergers commence around 
$z \sim 10$.  In this first series of mergers there are 
no supermassive BHs and mass ratios typically are from 
$0.06 \ltsim M_2/M_1 \ltsim 0.9$ 
(e.g. Figure~\ref{fig:M1_MR_z}).
The first series of mergers occurs in systems that are roughly 
$\Mvir \sim 100 \Mcrit \sim 10^{10} \hMsun$.
For halos described by the NFW density profile, the 
escape velocity from the halo center $\Vesc$, is related 
to the maximum circular velocity attained by the density 
profile $\Vmax$, by $\Vesc \simeq 3\Vmax$.  Typical 
values for this mass and redshift range are 
$\Vmax \sim 75 \kms$ so that $\Vesc \sim 225 \kms$.  
Thus, in most halos of interest, the velocity kick 
can likely be a significant fraction of the halo escape 
velocity or more, implying that the BH remnants are 
ejected from the halo centers.  

We estimate the potential impact of the gravitational 
radiation recoils on our results in the following way.  
For each merging pair, 
we assign the remnant a recoil speed set by the 
{\em upper limit} in Equation~(\ref{eq:vrecoil}).  For 
nearly $\sim 70\%$ of our BH mergers, recoils of this 
magnitude allow the remnant to escape from the halo.  
We cease to track escaped remnants.  Even for the remainder, 
the velocities are typically large enough that the ejected 
remnants make excursions that are a large fraction of 
the scale radius of the host halo profile ($\gtsim 0.3\rs$).
In this case it may be possible for some mechanism 
to return the remnants to the centers of halos rapidly, 
where they may experience further mergers; 
however, our simple model contains no such mechanism.  
In effect, the recoils cause our seed BHs to merge once 
with further merger activity effectively shut off.  
Therefore, this approach results in a minimum 
expected gravity wave signal from seed BHs formed according 
to the KBD mechanism by allowing each black hole to merger once 
at most. 

We note that for recoils of this size and in the 
absence of an efficient mechanism to return merger 
remnants to their halo centers, SBHs cannot be built up 
efficiently through mergers.  Instead, this case of 
a strong gravity wave recoil results in a population 
of ``orphaned'' BHs wandering in the 
intergalactic medium, similar to the 
study of \citet{VHM03}.

In our fiducial models, we neglect the effect of three-body 
interactions on merging black holes.  This occurs when a 
third BH falls into the coalescing BH binary at the halo 
center prior to completion of the BH merger.  The general 
effect of such interactions is that the lowest-mass BH of the 
three is ejected from the system with the consequence that 
the remaining binary is left more tightly bound than before 
\citep{hills_fullerton80}.  We estimate the importance of 
potential three-body interactions in the following way.  
We trace through our merger trees and determine the fraction of 
events for which a third BH falls within the inspiral radius 
of an existing binary before the existing binary has had time 
to merge according to 
$t_{\mathrm{inspiral}}$ in Eq.~(\ref{eq:time_in}).  

For our fiducial model, we find that this occurs in 
only $\sim 0.4 \%$ of BH mergers, so neglecting this 
effect is a reasonable approximation.  
We note that our estimates are in 
broad agreement with those of \citet{islam_etal04c}, 
which is not surprising because our model of halo evolution is 
similar to the model that they employ.  
However, \citet{VHM03} found three-body interactions to be 
non-negligible in their analysis. 
The discrepancy is likely due to the fact that 
\citet{VHM03} do not account for tidal mass 
loss from subhalos, thus greatly decreasing the dynamical 
friction infall timescale of a subhalo. In addition, 
a secondary effect is the fact that 
higher order subhalos in the merging tree may be stripped from 
infalling subhalos due to the tidal field of the main host 
system \citep[see][for a complete discussion]{ZETAL04}, an 
effect not taken into account in \citet{VHM03}.  
Even in the presence of three-body interactions, we expect that our model 
which includes maximal velocity recoils, should represent a lower bound on 
the predicted gravity wave background, because it allows each black hole 
to merge, at most, once.

\subsection{The Gravity Wave Background from Black Hole Mergers}
\label{sub:gw_bkgd}

We calculate the energy spectra for each merger as described 
previously in Section~\ref{sub:gw_binary}.  
For each potential pair, we calculate 
$\fmerger$ at the source 
(which depends only on the BH masses).  
Then, using our adopted values of 
$\tmig$ and $\beta$, we compute the well-understood time 
evolution of the emission frequency during inspiral using 
Eq.~(\ref{eq:time_in}) and determine the observed values 
of $\fmin$ and $\fmerger$.  We then estimate the duration of 
the merger and the observed value of $\fqnr$.  
The relations in the previous subsection then yield 
the energy spectrum for each merger event.  We 
present results for the total spectrum as well as the 
signals from inspiral, merger, and ringdown and decay 
which we treat together.  We now describe how we compute 
the gravity wave background from these spectra.

We compute the total gravity wave background using  
our Monte Carlo merger tree realizations.
Averaging over the merger tree realizations at a fixed final 
host mass allows us to compute the mean energy spectrum 
due to all BH pairs emitting gravitational radiation in a 
redshift range $dz$ about $z$, 
associated with progenitors of a halo of mass 
$\Mvir$ today.  
We denote this quantity $d \Egw(z,\Mvir) /dz d \ln f_s$, 
where $f_s = f (1 + z )$ is the frequency at the source.  
The levels of stochastic gravity wave backgrounds are 
often quoted in terms of the energy density in gravity 
waves per logarithmic frequency interval measured in units 
of the critical energy density of the universe, $\Omegagw(f)$.
We compute the comoving energy density in 
gravity waves by integrating $d \Egw/dz d\ln f_s$ 
over redshift and halo mass, 
\begin{eqnarray}
\label{eq:ogw}
\rhocrit c^2 \Omegagw(f) &=& \int \int \frac{dn(\Mvir,z=0)}{d \ln \Mvir}
\nonumber \\ 
&\times& 
\frac{d \Egw(z,\Mvir)}{dz d \ln f_s} (1+z)^{-1} dz \,  d\ln \Mvir
\end{eqnarray}

In Equation~(\ref{eq:ogw}), 
$dn(\Mvir,z=0)/d\ln(\Mvir)$ is the comoving number density of 
halos per logarithmic mass interval evaluated at redshift $z=0$.  
We assume that this function takes the form proposed by 
\citet{ST99}.  Formally, 
the integral over $\ln \Mvir$ should be taken from 
$\ln \Mvir = [-\infty,\infty]$.  In practice, we compute the 
integrand at discrete values of $\Mvir$ as described in 
Section~\ref{sub:haloevol}.  We then compute the integral 
using a power-law interpolation of 
$d \Egw(z,\Mvir)/dz d\ln f_s$ to evaluate the integrand of 
Eq.~(\ref{eq:ogw}) between neighboring bins of $\Mvir$.  
Once the energy density of the gravitational 
wave background is known, 
we can relate this back to the 
characteristic strain by 
\begin{eqnarray}
\label{eq:hcofogw}
h_c^2(f) & = & \frac{4G\rhocrit}{\pi f^2} \Omegagw(f)\mathrm{,}
\end{eqnarray}
so that 
\begin{eqnarray}
\label{eq:charstrain}
h_c(f) & \simeq & 8.8 \times 10^{-19} \Omegagw^{1/2}(f) 
\Bigg(\frac{\si}{f}\Bigg)\mathrm{.}
\end{eqnarray}

In the Section~\ref{section:results}, we present our results for the 
gravity wave background due to mergers between seed 
BHs in terms of the $h_c(f)$.

%
%
\subsection{Is The Gravity Wave Background Resolvable or Stochastic?}
\label{sub:stochasticity}
%
%

The background of gravity waves that we compute in 
Section~\ref{sub:gw_bkgd} is a sum of contributions from  
numerous individual mergers.  
It is important to determine whether the background 
of gravity waves that any merger scenario predicts 
can be resolved into discrete events.  In this section, 
we outline our method for determining whether our predicted 
signals are stochastic in nature, or resolvable.

Generally speaking, the gravity wave signal from a 
particular event can be considered ``resolved'' if it 
is detected above some {\it rms} signal-to-noise 
ratio, $\SN$.  This signal-to-noise ratio describes the 
likelihood of misinterpreting fluctuations in the instrumental 
noise or any known backgrounds that might 
be included in the calculation as the 
signal from the gravity wave source.  
In the case of a gravitational wave detector, 
the instrumental noise is quantified by 
the noise spectrum of the particular instrument 
\citep[e.g., see][]{thorne87,thorne96,larson_etal01}.  
Given a population of sources, 
we can estimate whether these sources 
will be resolved or not, 
by counting the number of sources that 
are above a certain $\SN$ in a given frequency 
bin $\Delta f$.  The frequency 
resolution is set 
by the length of the observation time 
$\tau_{\rm obs}$, as $\Delta f = \tau^{-1}_{\rm obs}$.  

Recently, \citet{C03} showed that a realistic 
criterion for a background signal, 
built from a sum of a number of individual events or 
sources, to be stochastic and unresolvable 
into individual sources is when there is 
more than one gravity wave source above a 
particular $\SN$ per $\sim$~eight 
frequency resolution bins, 
the so-called {\it eight-bin rule}.  
In what follows we use the 
eight-bin rule to determine whether our models 
predict resolvable or stochastic gravity wave 
backgrounds.  

For a gravity wave source inducing a characteristic strain $h_{c}(f)$, 
the signal-to-noise ratio at a frequency $f$, averaged over the 
orientation of the detector is given by 
\citep[see][for a derivation]{FH98}, 
\begin{eqnarray}
\label{eq:SNinspiralmerger}
\SN_{\Delta f} &=& \sqrt{ \int_f^{f + \Delta f } 
\frac{h^2_c(f)}{h^2_{\rm noise}(f)} \, \frac{df}{f} } \nonumber \\
 &=& 3.24 \times 10^{-27} \frac{(1+z)}{d_{\rm L}(z)} 
\sqrt{ \int_f^{f + \Delta f } \frac{dE/df_s}{h^2_{\rm noise}(f)} \frac{df}{f}}.
\end{eqnarray}
In Equation~(\ref{eq:SNinspiralmerger}), 
the luminosity distance to the source is in units of Mpc, and 
$dE/df_s$, the energy spectrum at the source, is 
in units of $\Msun {\rm m}^2 {\rm s}^{-1}$. 
The quantity $h_{\rm noise}(f)$ is the rms noise of the detector 
for sources at random directions and orientations.  
We assume a LISA sensitivity curve appropriate 
for an exposure time of $3$~years.  In 
this case, the frequency resolution bin has a 
width of $\Delta f \approx 10^{-8} {\rm s}^{-1}$. 

We estimate the number of sources per eight 
frequency resolution bins in the 
following way.  We first calculate the 
mean number of events per redshift 
interval from $z$ to $z + dz$, contributing above a 
specified $\SN$, in a frequency bin, 
for each bin halo mass in our merger 
tree realizations.  
We denote this quantity as 
$d \bar{N}(\Mvir,z)/dz$.  
We then calculate the mean number of 
events per comoving volume and time as 
\begin{equation}
\label{eq:dNdvdt}
\frac{d \bar{N}}{dVdt} = \int_0^\infty \frac{d \bar{N}(\Mvir,z)}{dz} 
\Bigg\vert  \frac{dz}{dt}  \Bigg\vert 
\frac{dn(\Mvir,z=0)}{d \Mvir dV} d\Mvir. 
\end{equation} 
Here, $dz/dt$ is given by $dz/dt = - (1+z)H_0 \sqrt{E(z)}$, where 
$E(z) = (1+z)^2 ( 1 + \Omega_M z ) - z ( 2 + z ) \Omega_\Lambda$
and $H_0$ is the present value of the Hubble constant.  
We evaluate the integral using the same 
power-law interpolation scheme that we used to 
evaluate the integral in Eq.~\ref{eq:ogw} above.  
From this quantity, we then estimate the total number of events 
observed per observed time as
\begin{equation}
\label{eq:Npert}
\frac{d \bar{N}}{dt_{\rm obs}} = \int_0^\infty \frac{d \bar{N}}{dVdt}  \frac{dV}
{dz} \frac{dz}{1+z}.
\end{equation}
The total number of observed events in an observation time $\tau$ 
is then simply 
\begin{equation}
\label{eq:Ntotal}
N_{\rm total} = \int_0^{\tau} \frac{d \bar{N}}{dt_{\rm obs}} dt_{\rm obs} \, .
\end{equation}
In Eq.~\ref{eq:Npert} the quantity $dV/dz$ is the comoving volume 
in a redshift interval between $z$ and $z+dz$ 
and is given by 
\begin{equation}
\label{eq:comovingvolume}
\frac{dV}{dz} = 4 \pi \Bigg( \frac{c}{H_0} \Bigg)^3 
\frac{1}{\sqrt{E(z)}} 
\Bigg( \int_0^z \frac{1}{\sqrt{E(z')}} {\rm d}z' \Bigg)^2 \, .
\end{equation}
%
%

\section{RESULTS}
\label{section:results}

\subsection{Stochasticity}
\label{sub:stochastic_results}
%
\begin{figure}[t]
\begin{center}
\includegraphics[height=7.25cm]{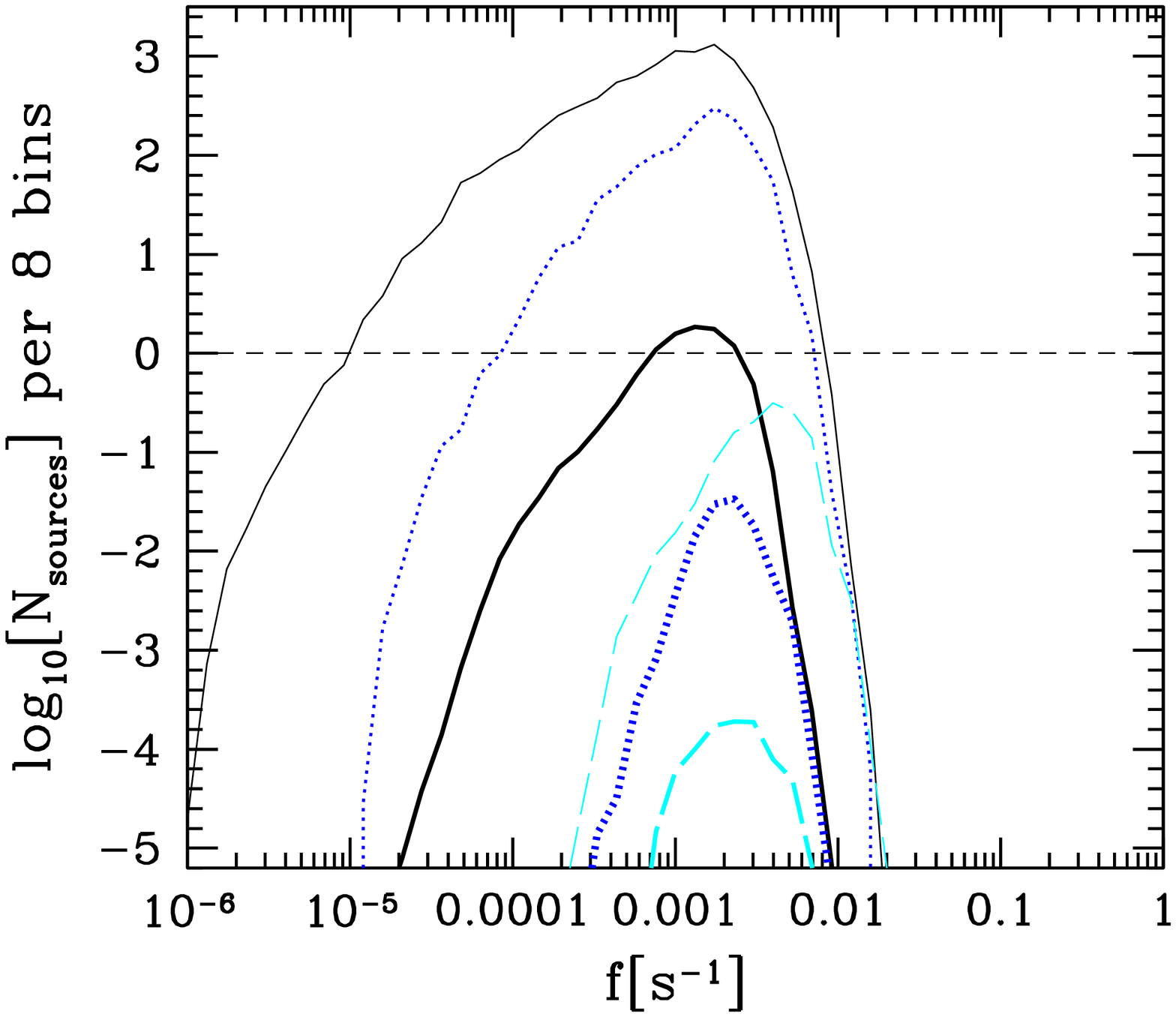}
\caption{
The number of black holes per eight frequency resolution 
bins for a 3 year observation time with LISA. 
{\em Thick} lines correspond to the number of sources with 
$\SN \ge 5$, while {\em thin} lines represent the number of 
sources with $\SN \ge 1$. 
{\em Solid lines} depict the number of sources in the fiducial 
case where the migration timescale is $\tmig = 10 Myr$, 
$\beta = 10^{-1}$ and reionization occurs at $\zre=12$. The 
{\em dotted line} corresponds to the same set of parameters 
with the inclusion of the effect of radiation recoil described 
in \S~\ref{sub:kicks}. The 
{\em short, dashed line} corresponds to the extreme scenario
where radiation recoil is included, the redshift of reionization 
is set at $\zre=16$ and the seed black hole masses are reduced by 
two orders of magnitude from the fiducial KBD mass (see 
\S~\ref{sub:background}).}
\label{fig:Nbins}
\end{center}
\end{figure}
%
Our first result addresses the issue of the stochasticity 
of the gravity wave background.  As described in 
\S~\ref{sub:stochasticity}, we calculate the number of 
individual black hole merger events that are above 
a given signal-to-noise ratio and assess stochasticity 
according to the eight bin rule.  
Figure~\ref{fig:Nbins} shows this result for the 
different variants of the KBD model that we study. 
Clearly, for a threshold signal-to-noise ratio of $\SN=1$, 
the number of sources per eight frequency bins in our 
fiducial scenario is quite large in 
the frequency range $10^{-2} \si \gtsim f \gtsim 10^{-5} \si$. 
This implies that the individual mergers will generally not 
be resolvable and that the gravity wave background 
is stochastic in nature over this frequency range.  
However, at a threshold signal-to-noise ratio of $\SN=5$, 
the gravity wave background becomes resolvable after 3 years 
of observation. 
In this case, the background is stochastic only in the small 
frequency range 
$2 \times 10^{-3}\si \gtsim f \gtsim 7 \times 10^{-4}\si$.  
Therefore, for low detection thresholds, the gravity wave 
background from a KBD-like model is stochastic, while for 
high detection thresholds it will be resolvable into 
individual merger events. 

The reason for the large number of sources 
in this scenario is easily understood.  
In models with large seeds, 
most pairs of black holes 
contribute a substantial gravity wave signal 
in the frequency range probed by an instrument like LISA.  
The gravity wave signal in high-mass seed black hole models 
is dominated by mergers between seeds just after formation, 
when the merger rate of halos and black holes is high 
(see Fig.~\ref{fig:N_z} and Fig.~\ref{fig:M1_MR_z}) 
and we will return to this point in the following section.  
Models based on the MR01-type scenario predict that the 
gravity wave background will be resolvable, even at a 
relatively low signal-to-noise ratio threshold of 
$\SN \sim 1$.  The reason is because only a 
relatively smaller number of black hole pairs 
in that scenario emit gravity waves in the LISA 
frequency range with large amplitudes.  In models with 
low-mass seeds, supermassive black holes grow predominantly 
through accretion, and do not reach masses that 
contribute significantly to the LISA frequency range 
until relatively low redshift. 

\subsection{The Gravitational Radiation Background}
\label{sub:background}

\begin{figure*}
\begin{center}
\includegraphics[height=13cm]{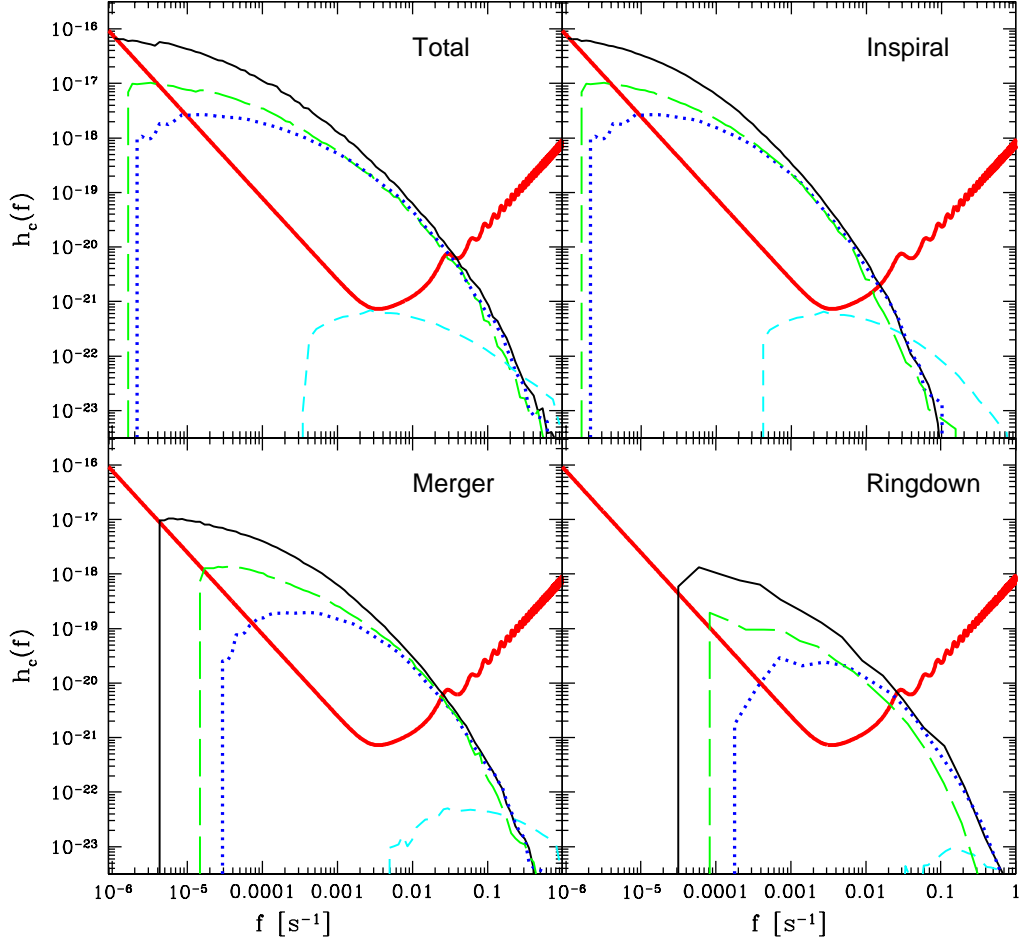}
\caption{ \small
The spectrum of the gravity wave background 
induced by the hierarchical merging 
of massive BH seeds.  
The {\em upper, left} panel shows the 
total gravity wave background while the 
other three panels show the background 
decomposed into contributions from mergers 
in the {\em inspiral}, {\em ringdown and decay}, 
and {\em merger} phases 
(proceeding clockwise from the upper left).  
In each panel, the {\em thick, solid} line that 
rises steeply at high frequency 
represents the LISA sensitivity to gravity wave bursts  
\citep{thorne87,thorne96,larson_etal01} 
computed using the LISA 
{\em online sensitivity curve generator}.  
The {\em thin, solid} lines correspond to 
the gravity wave background in our fiducial 
model with $\zre = 12$, $\tmig = 10$~Myr, 
and $\beta = 10^{-1}$.  
The {\em dotted} lines correspond to the same 
set of parameters with the inclusion of the 
effect of radiation recoil described in 
\S~\ref{sub:kicks}.  The {\em long-dashed} line 
corresponds to the same choice of parameters 
except for the redshift of reionization, which 
is set to $\zre = 16$, and with no radiation recoil.  
The {\rm short-dashed} line corresponds to an 
extreme scenario with maximum radiation recoil 
included, a redshift of reionization set at $\zre = 16$, 
and the seed BH masses reduced {\em artificially} 
by two orders of magnitude from the fiducial 
KBD mass scale.
} 
\label{fig:spectrum_split}
\end{center}
\end{figure*}

In this subsection, we present our estimates of the 
gravitational wave background produced by the merging 
of high-mass seed BHs formed at high redshift.  
Our primary result regarding the 
gravitational wave background from BH mergers 
is shown in Figure~\ref{fig:spectrum_split}.
This Figure shows the full gravity wave spectrum 
from BH mergers as well as a decomposition of the 
gravity wave spectrum into contributions from the 
different phases of BH coalescence and merger.  
This decomposition may be useful, for example, 
because the inspiral phase is understood significantly 
better than the subsequent phases of the merger, 
where numerical relativity is needed.  We also show 
the variation of our predictions with the two 
parameters that our model is most sensitive to, 
the redshift of reionization $\zre$, and the degree 
importance of radiation recoil after BH mergers.  
We present the background in terms of the characteristic 
strain induced as in Eq.~\ref{eq:charstrain}. 
In each panel of Fig.~\ref{fig:spectrum_split}, we 
also show the LISA sensitivity to bursts 
computed using noise 
spectral densities from the LISA 
{\em online sensitivity curve generator}) 
after a three year exposure 
\cite[][]{thorne87,thorne96,larson_etal01}.  

It is apparent from the results in Fig.~\ref{fig:spectrum_split}, 
that the background that we predict 
is considerable compared to the expected LISA instrumental 
sensitivity and that, if this model is correct, LISA should 
be able to measure this spectrum at very high signal-to-noise 
ratios.  Limits on pulsar timing residuals constrain the energy 
density in a stochastic gravitational wave background 
to roughly $\Omega_{\mathrm{gw}}(f) \ltsim 4 \times 10^{-9}$ 
in the frequency range $10^{-9} \ltsim (f/\si) \ltsim 10^{-8}$ 
\citep{lommen02}.  Our models are not constrained by 
these bounds because the spectra that we predict peak around 
typical frequencies of $f \sim 10^{-4}-10^{-2} \si$ and fall 
of rapidly on either side of the peak.  

As we describe in more detail below, 
varying the unknown parameters in the calculation 
within reasonable ranges does not change our basic result.  
We are left with the promising result that LISA should be 
able to confirm or essentially to rule out classes 
of models based on the formation of massive seed BHs 
because a considerable gravity wave 
background is one of the generic predictions of 
such a scenario.  LISA should be able to observe 
the relatively simple, well-understood phase of BH 
binary inspiral over more than $3$ orders of 
magnitude in frequency.  

Most previous studies employed significant growth of 
central BHs through accretion \citep{SETAL04a,SETAL04b}.  
In our model we do not include such a prescription, for two 
reasons.  First, we would like to examine the robustness 
of the KBD model and its predicted gravitational wave 
spectrum.  The spectrum would only be boosted by additional 
accretion onto central BHs.  Secondly, current 
prescriptions for accretion onto BHs 
that grow via mergers are uncertain and must be 
tuned in order to reproduce observed correlations 
at $z=0$ \cite[e.g.][]{VHM03}.  
Nevertheless, specific prescriptions for accretion 
can easily be incorporated into the framework of our 
calculation should growth through accretion become 
more well understood.

The general shape of the gravity wave spectrum 
can be understood in the following way.  
As the frequency of emission depends on the 
mass of the pair, the high-frequency regime is dominated by 
the mergers of relatively low-mass BHs with the original 
seed BH mass, while the low-frequency 
regime originates from the merger of more massive 
pairs.  The spectrum drops steeply at the 
high-frequency end for two reasons.  First, the 
spectra from individual mergers drop sharply at 
a maximal frequency set by the mass scale of the 
merging black holes, in this case the mass scale 
of the KBD seed BHs (see \S~\ref{sub:gw_binary}).  
Second, due to the hierarchical build-up of SBHs that 
we describe in this study, small masses contribute 
primarily at higher redshifts than high-mass BH 
pairs (see Figure \ref{fig:M1_MR_z}).  
The frequency of emitted radiation by 
small mass BHs experiences a larger 
redshift toward lower frequencies than 
that from corresponding pairs of larger mass BHs, 
resulting in a steepening of the gravity wave 
spectrum at high frequencies. 
The apparent cutoff that is seen at low 
frequencies basically reflects an upper limit on the 
resulting mass of BHs at $z=0$.  If the BH masses 
at $z=0$ are the result of merging of smaller mass BHs 
as we assume in this study, then this mass is directly proportional 
to the mass of seed BHs as well as the number of 
mergers that occurred in the process of forming the BH 
at $z=0$.  The sharpness of the cutoff results from our choice of 
a fixed value of the parameter $\beta= \fmin / \fmerger$. If instead 
we were able to assign a different value of $\beta$ for each 
one of the coalescing pairs, then the decline at lower frequencies 
would be smooth.  Our choice of a fixed value of 
$\beta$ for all pairs simplifies the calcualtion but 
does not change the essential results (see below).

We emphasize that our results are robust 
to variations of the migration 
timescale $\tmig$.  We can understand this directly 
from Fig.~\ref{fig:N_z} and Fig.~\ref{fig:M1_MR_z}.   
These Figures show that most of the potential 
BH pairs in this model are formed quite 
early on ($z \gtsim 10$) and, moreover, 
that many of these mergers are between the 
near-equal-mass seed BHs.  Thus, most pairs 
have of order $\sim$~many~Gyr to evolve 
toward inspiral and begin emitting appreciable 
gravity waves.  Varying $\tmig$ from $1$~Myr to $1$~Gyr 
results in only a small reduction in the 
gravitational wave background.  Only pairs that 
merge at redshifts $z \ltsim 0.1$ are excluded by 
this criterion and the number of these pairs is comparably 
quite small, as can be gleaned from Fig.~\ref{fig:N_z}.  
The only way to reduce significantly the signal would be 
to require migration times in excess of a Hubble time.  

Similarly, lowering the minimum frequency 
(or equivalently, increasing the maximum binary separation) 
at which gravitational radiation becomes the primary 
channel for energy loss, parameterized by 
$\beta \equiv \fmin/\fmerger$, allows BH binaries 
to spend more time evolving through the inspiral 
phase, boosting the low-frequency signal, which 
is dominated by inspiral, and reducing the 
high-frequency signal, where the subsequent 
phases of merger, ringdown, and decay dominate.  
Varying the value of $\beta$ from $10^{-1}$ to 
$10^{-3}$ results in, at maximum, an increase of $h_c(f)$ 
by a factor of $\sim 3$ in the low-frequency regime. 
On the other hand, increasing the value of 
$\beta$ up to $ 5 \times 10^{-1}$, the decrease 
in the characteristic strain is at most a factor of $2$.  
Therefore, varying the values of $\tmig$ and $\beta$ within 
their expected ranges does not significantly alter the 
main result of this study.

As we have already mentioned, unlike the parameters 
$\beta$ and $\tmig$, the gravity wave spectrum {\em is} 
sensitive to the redshift of reionization, $\zre$ 
(or any other cosmological epoch that effectively 
terminates the KBD process).  
In our fiducial model, we follow KBD and adopt a 
value of $\zre = 12$, which is shown by the thin, 
solid lines Fig.~\ref{fig:spectrum_split}.  
Taking a higher reionization redshift, as indicated by the 
cosmic microwave background polarization anisotropy 
\citep{spergel_etal03,page_etal03}, significantly reduces the 
low-frequency signal.  The dashed lines in 
Fig.~\ref{fig:spectrum_split} show the 
results for the KBD model with $\zre = 16$. 
In the frequency range of interest, the strain spectrum in 
this case is decreased by at most an order of magnitude 
relative to the fiducial case.  Notice that in the 
frequency range at which LISA is most sensitive 
($10^{-3} \ltsim (f/\si) \ltsim 10^{-2}$), 
the relative decrease is much smaller.  
The dominant reason for the reduction in the background 
at low frequency is that there are considerably 
fewer seed BHs in a model where the 
KBD mechanism is cut off at this 
high a redshift (see \S~\ref{sub:pophalos}).

Most previous studies have assumed that central BHs 
merge, one after the other, with infalling BHs.  
\citet{favata_etal04} point out that this may not 
necessarily hold because the gravity waves emitted 
during the merger are not emitted isotropically and 
may induce a recoil on the center of mass 
of the system that can be large compared 
to the escape velocities from typical dark matter halos.  
In this case, BHs can continue to merge {\em only} if 
some mechanism returns them to the centers of halos.  
We estimated the affect of this recoil on our 
predicted spectrum by allowing BHs to recoil with 
the {\em maximum} recoil velocity estimated in 
\citet{favata_etal04}.  
In this scenario, seed BHs effectively merge once and 
are then cut off from further mergers.

At the low frequency end (see Fig.~\ref{fig:spectrum_split}), 
the spectrum in the recoil scenario is roughly 
an order of magnitude lower than in our 
fiducial model; however, the reduction is significantly 
less at higher frequencies, particularly at the 
trough in the LISA noise spectrum.  The signal should 
still be detectable at high signal-to-noise by LISA 
at frequencies in excess of a few $\times 10^{-5}\ \si$.  
Again, the reason that this model is robust to the 
recoils is simple.  
In this seed BH formation scenario, BHs start with 
very large masses ($\sim 10^5 \Msun$) compared 
to the MR01-motivated models ($\mbh \sim 10^2 \Msun$).  
Mergers between these seeds already produce a sizable 
strain compared to the LISA noise curve.  
Therefore, even if the aggregation of BHs is cut-off 
after a single merger due to the recoil, the 
gravity wave background may still be quite large in the 
frequency regime that LISA will probe most reliably.

Finally, we present results for an extreme scenario that 
represents the minimal signal that can result from the type 
of BH formation model that we study here.  
In this extreme case, we include the maximum velocity recoil and 
the redshift of reionization is set to $\zre = 16$.  
In addition, we assume that the 
initial mass spectrum of BHs (see Fig.\ref{fig:bhi}) 
is shifted to smaller mass by a factor of $10^{-2}$.   
In this case, the characteristic BH mass is of order 
$\mbh \sim 10^3 \Msun$.  
In our estimate, we make this shift artificially, simply 
by resetting the KBD masses, but physically such a shift may 
be due to some unknown efficiency parameter 
associated with the formation of very massive BHs. 
In the case of the KBD model, seeds of this low mass would 
primarily form if only $\sim 1 \%$ of the mass that loses 
its angular momentum within $\sim 10$~Myr actually collapses 
into a black hole.  
The resultant gravity wave signal from this extreme model 
is represented by the short-dashed lines in 
Fig.~\ref{fig:spectrum_split}.  
The gravity wave background is significantly smaller in this model.  
Notice that because the BHs are smaller and there is minimal growth 
of BH mass via mergers, the low-frequency signal is essentially 
gone. However, if supermassive black holes are formed from 
such an extreme case of the KBD model, the dominant mode of 
SBH growth must be through accretion. 

In general, we note that 
the backgrounds presented here may still compete with other 
hypothetical backgrounds associated with, 
for example, galactic and extragalactic white dwarf 
binaries \citep{NETAL01,FF03}, the formation of the first generation 
of stars \citep{schneider_etal00}, supernovae \citep{buonanno_etal05} 
or excitations of 
additional degrees of freedom in theories of spacetimes 
with large extra dimensions \citep{hogan00},

\subsection{Number of resolvable events}
\label{sub:numresolvable}

As we showed in section \S~\ref{sub:stochasticity}, the gravity 
wave background will be resolved into individual sources with 
a signal-to-noise ratio of greater than 5 for most of the 
LISA frequency range. We can therefore estimate the number of 
sources that will be resolved in a given observation time. We assume 
as before that this time is 3 years, and we estimate the number of 
sources that are resolved in each one of the phases of gravity 
wave emission for each pair. We do this 
in the following way. For each potential 
pair, we estimate the signal-to-noise ratio for the inspiral and 
merger phases using 
Eq.~(\ref{eq:SNinspiralmerger}).  
For the inspiral phase, we define the frequency interval $\Delta f$, 
as the difference between $\fmerger$ and the frequency that 
the binary was emitting, 3 years 
(the observation time) before reaching $\fmerger$ (estimated 
using Eq.~(\ref{eq:time_in})).
We perform the integral over frequency in 
Eq.~(\ref{eq:SNinspiralmerger}) and determine whether the 
signal-to-noise ratio meets the requirement $\SN \ge 5$. 
For the merger phase, we perform the integral 
in Eq.~(\ref{eq:SNinspiralmerger}) with $\Delta f =\fqnr - \fmerger$.  
In the case of the ringdown phase, we approximate the energy spectrum 
as a $\delta$-function at the ringdown frequency, 
following the analysis of \citet{FH98}. In 
this case the signal-to-noise ratio is given by 
\begin{eqnarray} 
\SN_{\rm ringdown} &=& 3 \times 10^{-13} (1+z_{\rm qnr}) \nonumber \\
&\times& \frac{ 
\sqrt{ \epsilonr \fqnr
\xi (\mu, M_1, M_2 ) M_{\rm total,6}^3 }}{ f(a) \, d_L(z) \, h_{\rm noise} 
[\fqnr / (1+z_{\rm qnr}) ] }
\end{eqnarray}
where $f(a) = 1-0.63(1-a)^{3/10}$, and $a$ is as usual the 
dimensionless spin parameter. 
We perform these $\SN$ estimates for each pair in all halo 
masses of our merger tree realizations. We then calculate the mean 
number of events per redshift interval, $d\bar{N}(\Mvir,z)/dz$, and 
follow the analogous sequence going from 
Eq.~(\ref{eq:dNdvdt}) to Eq.~(\ref{eq:Ntotal}) 
to get the total number of events with 
$\SN \ge 5$ in a 3 year of observation 
with LISA. 

We find that for our fiducial case with $\zre=12$, $\tmig=10 \, {\rm Myr}$ and 
$\beta=10^{-1}$ that in a 3 year integration, LISA should be able to 
measure $\sim 400$ inspiral, $\sim 2300$ merger and $\sim 1100$ ringdown 
events above a signal-to-noise ratio of $\SN \ge 5$. For the case with 
the same set of parameters but including also the effects of radiation 
recoil described in section \S~\ref{sub:kicks}, the corresponding number
of resolved events in the inspiral, merger and ringdown phases is 250, 
700 and 500 respectively, while for the extreme scenario with 
maximum radiation recoil, a redshift of reionization of $\zre = 16$ and 
seed BH masses which are two orders of magnitude lower from the fiducial 
KBD mass scale these numbers are 1, 60 and 0.1. 
In all cases,  the number of merger and  ringdown events is higher
than the number of inspiral events. The reason is due to the fact that
most  of the black  hole masses  involved (see  Fig.~\ref{fig:N_z} and
Fig.~\ref{fig:M1_MR_z}) are such that the signal-to-noise ratio during
merger and  ringdown is generally  larger than the  signal-to-noise of
inspiral.    There   are  two   reasons   for   this.   The   inspiral
signal-to-noise  ratio is  roughly a  factor $\sim  M_1 M_2  /  (M_1 +
M_2)^2$  smaller  than the  signal-to-noise  ratio  for merger  and/or
ringdown events and, in many cases, inspiral binaries do not enter the
LISA sensitivity  range until only few  weeks prior to  the merger and
subsequent ringdown phase\footnote{for detection schemes in this case,
see the extensive discussion in Section VI of \citet{FH98}.}.

As with the gravity wave spectrum, 
the numbers of resolved events reflect the underlying characteristics of 
the KBD model.  Massive seed black holes which are formed early experience 
a large number of mergers. In addition, the seed masses are such that 
the emitted gravity wave radiation is in the frequency range of a space 
detector such as LISA. In supermassive black hole formation 
scenarios which are based on the MR01 
model for seed BH formation, seed masses are generally small.  As such, 
the emitted radiation does not move into the LISA frequency range until 
relatively lower redshifts, at which time the merger rate of the halos 
that host black holes is significantly decreased.  
Thus, just as the gravity wave background, the number of 
resolved events in MR01 models is generally significantly 
smaller than what is found in KBD-type high-mass BH seed models.

\section{CONCLUSIONS}
\label{section:conclusions}

In this paper, we studied the theoretical 
gravitational wave background predicted by 
models where supermassive black holes form from 
a population of high mass seeds and models where 
mergers are an important channel of SBH growth.  
We studied as a particular example, 
the supermassive BH seed 
formation scenario of \citet{KBD04}.  
Our primary results can be summarized as follows.

\begin{itemize}

\item[1.] The KBD seed BH formation scenario makes a 
{\em generic and testable} prediction 
of a very large gravity wave background 
that should be resolvable at high signal-to-noise 
by a space gravity wave detector such as the proposed
LISA interferometer.  This prediction is 
insensitive to many of the details of the model.  
In addition, we have neglected any model of BH 
growth via accretion which would serve to increase 
the signal.  This has the important implication that 
instruments with LISA characteristics 
should be able to confirm or to rule out the 
KBD scenario and similar models with high mass 
BH seeds and that future gravity wave detectors 
should be able to help 
assess the importance of mergers in the formation 
of SBHs and the establishment of the observed correlations 
between the black hole mass and the properties of its 
host spheroid and halo.  Moreover, 
if the background is detected, an instrument like LISA may 
be able to study the spectrum in detail 
providing constraints on some of the 
uncertain aspects of the model.

\item[2.] In general, the detectability of the 
predicted gravitational wave background is not 
sensitive to the details of the mechanism of 
BH migration if BH seeds are formed at high redshift.  
As long as migration occurs on a timescale less 
than $H_0^{-1}$, gravity waves from the inspiral 
phase of the mergers should still be observable.

\item[3.] For KBD-like models with an initial mass 
function of BH seeds that has a typical mass scale of 
$\mbh \gtsim 10^3 \Msun$, the gravity wave signal is robust 
to shifts in the redshift of reionization and a very 
generous reduction in the number of mergers due 
to radiation recoil effects and three body interactions.

\item[4.] While the magnitude of the gravity wave background means that it
provides a means of testing BH formation mechanisms, such a signal would
also represent a significant background contaminant for any future
low-frequency, gravity wave observatories such as BBO or DECIGO
\citep{seto_etal01}.  The gravity wave signal due to models of
high-mass seeds would represent a potentially large stochastic background
because any such instrument would have a greater sensitivity than LISA.
This large background may then inhibit such detectors from achieving
their aims of detecting the stochastic gravity wave 
backgrounds produced during inflation \citep[e.g.][]{turner97}, from a
first generation of supernovae \citep[e.g.][]{buonanno_etal05}, or perhaps
from other nonstandard physics \citep[e.g.][]{hogan00}.

\item[5.] The modeling that we present here can be used 
to provide detailed predictions for the gravity wave 
signals from a wide class of models of BH formation and 
their subsequent growth (including growth via accretion) 
and merging.

\end{itemize}

A detection of a gravitational wave background 
will be an enormous discovery in and of itself.  
One important reason is that it will open up a new 
window with which to observe the high-redshift universe.  
We showed that such a detection may 
yield important information about the formation and 
subsequent evolution of supermassive BHs.  
The presence of a background similar to our predictions 
would suggest the presence of numerous merger events, 
providing indirect evidence for a high merger rate of massive 
BHs.  However, it is hard to disentangle the merger rate 
from any signal because of a degeneracy between the 
BH masses and the redshifts that gravitational waves are 
emitted \citep{H02} unless independent redshifts measurements 
can be made.  Though this may be possible for individual merger 
events, in the case of a stochastic background this remains an 
outstanding difficulty.  

Further information may be gleaned if spectral 
features are present in the spectrum.  In this calculation, 
we neglected the gravitational wave emission that may result 
from the initial collapse of the seed BHs.  
This could result in the presence of a 
broadened ``emission line'' at a frequency which would 
correspond roughly to the redshifted quasi-normal-ringing 
frequency of the initial BH seed, and constrain the mass of 
seed BHs (see \citet{FHH02} for a discussion of the 
gravitational wave emission from collapsing objects). 
If the background could be measured at high signal-to-noise, 
the broadening of the line would provide information on the 
redshift of seed BH formation as well as the spectrum of seed 
BH mass, while the amplitude of the feature would indicate 
the abundance of the host halos that harbor the seed BHs \citep{KZ05b}.

A non-detection of this predicted background would place very 
strong limits on the viability of the model proposed by 
\citet{KBD04} and similar models of SBH formation from high-mass 
seeds.  Most notably it would severely constrain the mass and 
abundance of seed BHs.  If this background were not detected, then 
BH seeds would be required to be significantly smaller than the 
seeds proposed by KBD ($\mbh \ltsim 10^3 \Msun$) or 
significantly scarcer.  
Lack of a detection would indicate that growth via 
mergers of numerous pairs is not the dominant growth 
mechanism of SBHs and that KBD seed BHs must 
be so rare that they play, at most, a very limited role in 
the formation of SBHs in the centers of galaxies.  
In this case, growth via accretion would have to be the 
dominant growth mechanism. 
If, in fact, this is the case, then there might be 
consequences for the spin evolution of BHs 
as was pointed out by \citet{VMQR04}.  

These points demonstrate how a gravity wave detector like LISA 
will open a fundamentally new window on the high-redshift universe, 
providing a wealth of information on the formation and growth 
of supermassive black holes and probing the physics of galaxy 
formation.  LISA will probe galaxy formation processes at very 
high redshift, a task not attainable by other forms of observation 
except for possibly 21~$\mathrm{cm}$ emission \citep[e.g.][]{cabron_etal04}. 
Supermassive BHs and their formation are directly linked 
to the formation of galaxies.  Information gained 
from the gravitational wave background induced from BH 
mergers will inevitably provide useful constraints on 
high-redshift galaxy formation and insight into the 
processes that set the observed correlations between 
supermassive BHs and their hosts.  
An instrument with the sensitivity and frequency range similar to 
LISA will be able to confirm, disprove, or 
distinguish between mechanisms of SBH formation that produce 
otherwise identical predictions at $z=0$ 
(a not uncommon situation as many models have tunable parameters).  
We demonstrated in this manuscript how LISA will be able to 
distinguish between the gravitational wave background induced 
by the KBD model seed BHs and the models 
that stem from the MR01 scenario.  However, the new observational 
window on the universe that space-based gravity wave detectors 
will provide promises many 
further constraints on the fundamental processes of galaxy 
formation at high redshift.

\begin{acknowledgments}
We thank the anonymous referee for numerous 
useful suggestions that improved the quality of this manuscript. 
We are pleased to thank John Beacom, James Bullock, 
Neal Dalal, Daniel Holz, Scott Hughes, 
Stelios Kazantzidis, Andrey Kravtsov,
Pantelis Papadopoulos, Eduardo Rozo, Gary Steigman, 
Louis Strigari, Frank van den Bosch, 
Terry Walker, Casey Watson, David Weinberg and 
Simone Weinmann for many helpful comments.  
We would like to thank Gianfranco Bertone for many useful 
conversations 
throughout the course of this work and Mark Knopfler for 
constant support. 
We acknowledge use of the 
{\em online sensitivity curve generator} for space-based 
gravitational wave observatories maintained by 
S.~L. Larson.  
We would like to thank the 2004 Santa Fe Cosmology Workshop 
where this work was initiated.  
This research made use of the NASA Astrophysics Data System. 
SMK was partially supported by Los Alamos National Laboratory (under  
DOE contract W-7405-ENG-36). 
ARZ is funded by The Kavli Institute for Cosmological Physics 
at The University of Chicago and The National Science Foundation 
through grant No. NSF PHY 0114422.  
\end{acknowledgments}


\begin{thebibliography}{117}
\expandafter\ifx\csname natexlab\endcsname\relax\def\natexlab#1{#1}\fi

\bibitem[{{Abel} {et~al.}(2000){Abel}, {Bryan}, \& {Norman}}]{ABN00}
{Abel}, T., {Bryan}, G., \& {Norman}, M. 2000, ApJ, 540, 39

\bibitem[{Adams {et~al.}(2003)Adams, Graff, Mbonye, \& Richstone}]{AGR03}
Adams, F.~C., Graff, D.~S., Mbonye, M., \& Richstone, D.~O. 2003, ApJ, 591, 125

\bibitem[{Adams {et~al.}(2001)Adams, Graff, \& Richstone}]{AGR01}
Adams, F.~C., Graff, D.~S., \& Richstone, D.~O. 2001, ApJ, 551, L31

\bibitem[{Armitage \& Natarajan(2002)}]{AN02}
Armitage, P.~J. \& Natarajan, P. 2002, ApJ, 567, L9

\bibitem[{Barafee {et~al.}(2001)Barafee, Heger, \& Woosley}]{BHW01}
Barafee, I., Heger, A., \& Woosley, S.~E. 2001, ApJ, 550, 890

\bibitem[{Begelman {et~al.}(1980)Begelman, Blandford, \& Rees}]{BBR80}
Begelman, M.~C., Blandford, R.~D., \& Rees, M.~J. 1980, Nature, 287, 307

\bibitem[{{Bond} {et~al.}(1991){Bond}, {Cole}, {Efstathiou}, \&
  {Kaiser}}]{bond_etal91}
{Bond}, J.~R., {Cole}, S., {Efstathiou}, G., \& {Kaiser}, N. 1991, ApJ, 379,
  440

\bibitem[{Brandt \& Seidel(1995)}]{BS95}
Brandt, S.~R. \& Seidel, E. 1995, \prd, 52, 870

\bibitem[{Bromm {et~al.}(2001)Bromm, Coppi, \& Larson}]{BCL99}
Bromm, V., Coppi, P., \& Larson, R.~B. 2001, ApJ, 527, L5

\bibitem[{{Bromm} \& {Loeb}(2003)}]{BL03}
{Bromm}, V. \& {Loeb}, A. 2003, ApJ, 596, 34

\bibitem[{Bullock {et~al.}(2001)Bullock, Dekel, Kolatt, Kravtsov, Klypin,
  Porciani, \& Primack}]{B01}
Bullock, J.~S., Dekel, A., Kolatt, T.~S., Kravtsov, A.~V., Klypin, A.~A.,
  Porciani, C., \& Primack, J.~R. 2001, Astrophys.J., 555, 240 (B01)

\bibitem[{{Buonanno} {et~al.}(2005){Buonanno}, {Sigl}, {Raffelt}, {Janka}, \&
  {M{\"{u}}ller}}]{buonanno_etal05}
{Buonanno}, A., {Sigl}, G., {Raffelt}, G.~G., {Janka}, H.-T., \&
  {M{\"{u}}ller}, E. 2005, PRD, Submitted (astro-ph/0412277)

\bibitem[{{Chandrasekhar}(1943)}]{chandrasekhar43}
{Chandrasekhar}, S. 1943, ApJ, 97, 255

\bibitem[{{Chen} {et~al.}(2003){Chen}, {Jing}, \& {Yoshikaw}}]{chen_etal03}
{Chen}, D.~N., {Jing}, Y.~P., \& {Yoshikaw}, K. 2003, \apj, 597, 35

\bibitem[{Cornish(2003)}]{C03}
Cornish, N.~J. 2003

\bibitem[{{Cornish} \& {Larson}(2001)}]{cornish_larson01}
{Cornish}, N.~J. \& {Larson}, S.~L. 2001, Class. Quant. Grav., 18, 3473

\bibitem[{Dekel \& Birnboim(2004)}]{DB04}
Dekel, A. \& Birnboim, Y. 2004, Submitted (astro-ph/0412300)

\bibitem[{Detweiler(1979)}]{D79}
Detweiler, S. 1979, In Sources of Gravitational Radiation (Cambridge: Cambridge
  University Press, 1979, Smarr, L. L., ed.)

\bibitem[{Di~Matteo {et~al.}(2005)Di~Matteo, Springel, \& Hernquist}]{DiMSH05}
Di~Matteo, T., Springel, V., \& Hernquist, L. 2005, \nat, 433, 604

\bibitem[{Ebisuzaki {et~al.}(2001)Ebisuzaki, Makino, Tsuru, Funato, Portegies,
  Hut, McMillan, Matsushita, Matsumoto, \& Kawabe}]{EETAL01}
Ebisuzaki, T., Makino, J., Tsuru, T., Funato, Y., Portegies, Z.~S., Hut, P.,
  McMillan, S., Matsushita, S., Matsumoto, H., \& Kawabe, R. 2001, ApJ, 552,
  L19

\bibitem[{Echeverria(1989)}]{E89}
Echeverria, F. 1989, Phys. Rev., D40, 3194

\bibitem[{Eisenstein \& Loeb(1995)}]{EL95}
Eisenstein, D.~J. \& Loeb, A. 1995, ApJ, 443, 11

\bibitem[{Enoki {et~al.}(2004)Enoki, Inoue, Nagashima, \& Sugiyama}]{EETAL04}
Enoki, M., Inoue, K.~T., Nagashima, M., \& Sugiyama, N. 2004, ApJ, 615, 19

\bibitem[{Enoki {et~al.}(2003)Enoki, Nagashima, \& Gouda}]{ENG03}
Enoki, M., Nagashima, M., \& Gouda, N. 2003, \pasj, 55, 133

\bibitem[{Farmer \& Phinney(2003)}]{FF03}
Farmer, A.~J. \& Phinney, E.~S. 2003, MNRAS, 346, 1197

\bibitem[{{Favata} {et~al.}(2004){Favata}, {Hughes}, \& {Holz}}]{favata_etal04}
{Favata}, M., {Hughes}, S.~A., \& {Holz}, D.~E. 2004, ApJL, 607, L5

\bibitem[{Ferrarese(2002)}]{F02}
Ferrarese, L. 2002, ApJ, 578, 90

\bibitem[{{Ferrarese} \& {Ford}(2005)}]{F04}
{Ferrarese}, L. \& {Ford}, H. 2005, Space Science Reviews, 116, 523

\bibitem[{Ferrarese \& Merritt(2000)}]{FM00}
Ferrarese, L. \& Merritt, D. 2000, ApJ, 539, L9

\bibitem[{{Fitchett}(1983)}]{fitchett83}
{Fitchett}, M.~J. 1983, MNRAS, 203, 1049

\bibitem[{{Flanagan} \& {Hughes}(1998a)}]{FH98}
{Flanagan}, E.~E. \& {Hughes}, S.~A. 1998a, \prd, 57, 4535 (FH98)

\bibitem[{{Flanagan} \& {Hughes}(1998b)}]{flanagan_hughes98b}
---. 1998b, \prd, 57, 4566

\bibitem[{Fryer {et~al.}(2002)Fryer, Holz, \& Hughes}]{FHH02}
Fryer, C.~L., Holz, D.~E., \& Hughes, S.~A. 2002, ApJ, 565, 430

\bibitem[{Fryer {et~al.}(2001)Fryer, Woosly, \& Heger}]{FWH01}
Fryer, C.~L., Woosly, S.~E., \& Heger, A. 2001, ApJ, 550, 372

\bibitem[{Gebhardt {et~al.}(2000)}]{GETAL00}
Gebhardt, K. {et~al.} 2000, ApJ, 539, L13

\bibitem[{Gnedin(2001)}]{G01}
Gnedin, O.~Y. 2001, Class. Quant. Grav., 18, 3983

\bibitem[{Gould \& Rix(2000)}]{GR00}
Gould, A. \& Rix, H. 2000, ApJ, 532, L29

\bibitem[{Granato {et~al.}(2001)}]{GLETAL99}
Granato, G.~L. {et~al.} 2001, MNRAS, 324, 757

\bibitem[{Haehnelt(1994)}]{H94}
Haehnelt, M.~G. 1994, MNRAS, 269, 199

\bibitem[{Haehnelt {et~al.}(1998)Haehnelt, Natarajan, \& Rees}]{HNR98}
Haehnelt, M.~G., Natarajan, P., \& Rees, M.~J. 1998, MNRAS, 300, 827

\bibitem[{Haehnelt \& Rees(1993)}]{HR93}
Haehnelt, M.~G. \& Rees, M.~J. 1993, MNRAS, 263, 168

\bibitem[{Haiman \& Menou(2000)}]{HM00}
Haiman, Z. \& Menou, K. 2000, ApJ, 531, 42

\bibitem[{Hatziminaoglou {et~al.}(2003)Hatziminaoglou, Mathez, Solanes,
  Manrique, \& Salvador-Sole}]{HMSMS03}
Hatziminaoglou, E., Mathez, G., Solanes, J.-M., Manrique, A., \& Salvador-Sole,
  E. 2003, MNRAS, 343, 692

\bibitem[{Heger \& Woosley(2002)}]{HW02}
Heger, A. \& Woosley, S.~E. 2002, ApJ, 567, 532

\bibitem[{{Hills} \& {Fullerton}(1980)}]{hills_fullerton80}
{Hills}, J.~G. \& {Fullerton}, L.~W. 1980, Astron. J., 85, 1281

\bibitem[{Hils \& Bender(1995)}]{HB95}
Hils, D. \& Bender, P.~L. 1995, ApJ, 445, L7

\bibitem[{{Hogan}(2000)}]{hogan00}
{Hogan}, C.~J. 2000, \prl, 85, 2044

\bibitem[{Hughes(2002)}]{H02}
Hughes, S. 2002, MNRAS, 331, 805

\bibitem[{Hughes \& Blandford(2003)}]{HR03}
Hughes, S.~A. \& Blandford, R.~D. 2003, ApJ, 585, L101

\bibitem[{{Islam} {et~al.}(2004a){Islam}, {Taylor}, \& {Silk}}]{islam_etal04a}
{Islam}, R.~R., {Taylor}, J.~E., \& {Silk}, J. 2004a, MNRAS, 354, 427

\bibitem[{{Islam} {et~al.}(2004b){Islam}, {Taylor}, \& {Silk}}]{islam_etal04b}
---. 2004b, MNRAS, 354, 443

\bibitem[{{Islam} {et~al.}(2004c){Islam}, {Taylor}, \& {Silk}}]{islam_etal04c}
---. 2004c, MNRAS, 354, 629

\bibitem[{{Jackson}(1975)}]{jackson75}
{Jackson}, J.~D. 1975, Classical Electrodynamics (92/12/31, New York: Wiley,
  1975, 2nd ed.)

\bibitem[{Kawaguchi {et~al.}(2004)Kawaguchi, Aoki, Ohta, \& Collin}]{KETAL04}
Kawaguchi, T., Aoki, K., Ohta, K., \& Collin, S. 2004, {\aap}, 420, L23

\bibitem[{{Kazantzidis} {et~al.}(2004){Kazantzidis}, {Mayer}, {Colpi}, {Madau},
  {Debattista}, {Moore}, {Wadsley}, {Stadel}, \& {Quinn}}]{kazantzidis_etal04}
{Kazantzidis}, S., {Mayer}, L., {Colpi}, M., {Madau}, P., {Debattista}, V.,
  {Moore}, B., {Wadsley}, J., {Stadel}, J., \& {Quinn}, T. 2004, ApJ, 623, L67

\bibitem[{{Klypin} {et~al.}(2002){Klypin}, {Zhao}, \&
  {Somerville}}]{klypin_etal04}
{Klypin}, A., {Zhao}, H., \& {Somerville}, R.~S. 2002, \apj, 573, 597

\bibitem[{Kormendy \& Richstone(1995)}]{KR95}
Kormendy, J. \& Richstone, D. 1995, ARA\&A, 33, 581

\bibitem[{Koushiappas {et~al.}(2004)Koushiappas, Bullock, \& Dekel}]{KBD04}
Koushiappas, S.~M., Bullock, J.~S., \& Dekel, A. 2004, MNRAS, 354, 292 (KBD)

\bibitem[{{Koushiappas} \& {Zentner}(2005)}]{KZ05b}
{Koushiappas}, S.~M. \& {Zentner}, A.~R. 2005, In preparation

\bibitem[{{Lacey} \& {Cole}(1993)}]{lacey_cole93}
{Lacey}, C. \& {Cole}, S. 1993, MNRAS, 262, 627

\bibitem[{{Larson} {et~al.}(2001){Larson}, {Hiscock}, \&
  {Hellings}}]{larson_etal01}
{Larson}, S.~L., {Hiscock}, W.~A., \& {Hellings}, R.~W. 2001, PRD, 62, 062001

\bibitem[{Lee(1995)}]{L95}
Lee, H.~M. 1995, MNRAS, 272, 605

\bibitem[{Lin \& Pringle(1987)}]{LP87}
Lin, D. N.~C. \& Pringle, J.~E. 1987, MNRAS, 225, 607

\bibitem[{Loeb \& Rasio(1994)}]{LR94}
Loeb, A. \& Rasio, F.~A. 1994, ApJ, 432, 52

\bibitem[{{Lommen}(2002)}]{lommen02}
{Lommen}, A.~N. 2002, in Neutron Stars, Pulsars, and Supernova Remnants, 114

\bibitem[{Madau \& Rees(2001)}]{MR01}
Madau, P. \& Rees, M.~J. 2001, ApJ, 551, L27 (MR01)

\bibitem[{Magorrian {et~al.}(1998)}]{M98}
Magorrian, J. {et~al.} 1998, AJ, 115, 2285

\bibitem[{Matsubayashi {et~al.}(2004)Matsubayashi, Shinkai, \& T.}]{MSE04}
Matsubayashi, T., Shinkai, H., \& T., E. 2004, ApJ, 614, 864

\bibitem[{Menou {et~al.}(2001)Menou, Haiman, \& Narayanan}]{MHN01}
Menou, K., Haiman, Z., \& Narayanan, V.~K. 2001, ApJ, 558, 535

\bibitem[{Merritt \& Ferrarese(2001{\natexlab{a}})}]{MF01a}
Merritt, D. \& Ferrarese, L. 2001{\natexlab{a}}, MNRAS, 320, L30

\bibitem[{Merritt \& Ferrarese(2001{\natexlab{b}})}]{MF01b}
---. 2001{\natexlab{b}}, ApJ, 547, 140

\bibitem[{Merritt \& Milosavljevic(2004)}]{MM04}
Merritt, D. \& Milosavljevic, M. 2004, To appear in Living Reviews of
  Relativity (astro-ph/0410364)

\bibitem[{Milosavljevi{\'{c}} \& Merritt(2001)}]{MM01}
Milosavljevi{\'{c}}, M. \& Merritt, D. 2001, ApJ, 563, 34

\bibitem[{Miralda-Escud{\'{e}} \& Kollmeier(2003)}]{MEK03}
Miralda-Escud{\'{e}}, J. \& Kollmeier, J.~A. 2003, ApJ, Submitted
  (astro-ph/0310717)

\bibitem[{{Misner} {et~al.}(1973){Misner}, {Thorne}, \&
  {Wheeler}}]{misner_etal73}
{Misner}, C.~W., {Thorne}, K.~S., \& {Wheeler}, J.~A. 1973, Gravitation (San
  Francisco: W.~H. Freeman \& Co.)

\bibitem[{Murray {et~al.}(2005)Murray, Quataert, \& Thompson}]{MQT04}
Murray, N., Quataert, E., \& Thompson, T.~A. 2005, \apj, 618, 569

\bibitem[{{Navarro} {et~al.}(1997){Navarro}, {Frenk}, \& {White}}]{nfw97}
{Navarro}, J.~F., {Frenk}, C.~S., \& {White}, S.~D.~M. 1997, ApJ, 490, 493
  (NFW)

\bibitem[{Nelemans {et~al.}(2001)Nelemans, Yungelson, \&
  Portegies-Zwart}]{NETAL01}
Nelemans, G., Yungelson, L.~R., \& Portegies-Zwart, S.~F. 2001, A\&A, 375, 890

\bibitem[{{Page} {et~al.}(2003){Page}, {Nolta}, {Barnes}, {Bennett}, {Halpern},
  {Hinshaw}, {Jarosik}, {Kogut}, {Limon}, {Meyer}, {Peiris}, {Spergel},
  {Tucker}, {Wollack}, \& {Wright}}]{page_etal03}
{Page}, L., {Nolta}, M.~R., {Barnes}, C., {Bennett}, C.~L., {Halpern}, M.,
  {Hinshaw}, G., {Jarosik}, N., {Kogut}, A., {Limon}, M., {Meyer}, S.~S.,
  {Peiris}, H.~V., {Spergel}, D.~N., {Tucker}, G.~S., {Wollack}, E., \&
  {Wright}, E.~L. 2003, \apjs, 148, 233

\bibitem[{Pizzolato \& Soker(2004)}]{PS04}
Pizzolato, F. \& Soker, N. 2004, MNRAS, Sumitted (astro-ph/0407042)

\bibitem[{Quinlan(1996)}]{Q96}
Quinlan, G.~D. 1996, NewA, 1, 35

\bibitem[{Quinlan \& Shapiro(1990)}]{QS90}
Quinlan, G.~D. \& Shapiro, S.~L. 1990, ApJ, 356, 483

\bibitem[{Rhook \& Wyithe(2005)}]{RW05}
Rhook, K.~J. \& Wyithe, J. S.~B. 2005, \mnras, Submitted (astro-ph/0503210)

\bibitem[{Salucci {et~al.}(1999)Salucci, Szuszkiewicz, Monaco, \&
  Danese}]{SSMD99}
Salucci, P., Szuszkiewicz, E., Monaco, P., \& Danese, L. 1999, MNRAS, 307, 637

\bibitem[{Sazonov {et~al.}(2004)Sazonov, Ostriker, Ciotti, \& Sunyaev}]{SOCS04}
Sazonov, S.~Y., Ostriker, J.~P., Ciotti, L., \& Sunyaev, R.~A. 2004, \mnras,
  Submitted (astro-ph/0411086)

\bibitem[{{Schneider} {et~al.}(2000){Schneider}, {Ferrara}, {Ciardi},
  {Ferrari}, \& {Matarrese}}]{schneider_etal00}
{Schneider}, R., {Ferrara}, A., {Ciardi}, B., {Ferrari}, V., \& {Matarrese}, S.
  2000, MNRAS, 317, 385

\bibitem[{Sesana {et~al.}(2004{\natexlab{a}})Sesana, Haardt, Madau, \&
  Volonteri}]{SETAL04a}
Sesana, A., Haardt, F., Madau, P., \& Volonteri, M. 2004{\natexlab{a}}, ApJ,
  611, 623

\bibitem[{Sesana {et~al.}(2004{\natexlab{b}})Sesana, Haardt, Madau, \&
  Volonteri}]{SETAL04b}
---. 2004{\natexlab{b}}, ApJ, 623, 23

\bibitem[{{Seto} {et~al.}(2001){Seto}, {Kawamura}, \& {Nakamura}}]{seto_etal01}
{Seto}, N., {Kawamura}, S., \& {Nakamura}, T. 2001, PRL, 87, 221103

\bibitem[{Shaerer(2002)}]{S02}
Shaerer, D. 2002, A\&A, 382, 28

\bibitem[{Shapiro \& Shibata(2002)}]{SS02}
Shapiro, S.~L. \& Shibata, M. 2002, ApJ, 577, 904

\bibitem[{Shapiro \& Teukolsky(1983)}]{ST83}
Shapiro, S.~L. \& Teukolsky, S.~A. 1983, Black holes, white dwarfs and neutron
  stars (Wiley)

\bibitem[{Sheth \& Tormen(1999)}]{ST99}
Sheth, R.~K. \& Tormen, G. 1999, MNRAS, 308, 119

\bibitem[{Smarr(1979)}]{S79}
Smarr, L.~L. 1979, In Sources of Gravitational Radiation (Cambridge: Cambridge
  University Press, 1979, Smarr, L. L., ed.)

\bibitem[{{Somerville} \& {Kolatt}(1999)}]{somerville_kolatt99}
{Somerville}, R.~S. \& {Kolatt}, T.~S. 1999, MNRAS, 305, 1

\bibitem[{{Spergel} {et~al.}(2003){Spergel}, {Verde}, {Peiris}, {Komatsu},
  {Nolta}, {Bennett}, {Halpern}, {Hinshaw}, {Jarosik}, {Kogut}, {Limon},
  {Meyer}, {Page}, {Tucker}, {Weiland}, {Wollack}, \&
  {Wright}}]{spergel_etal03}
{Spergel}, D.~N., {Verde}, L., {Peiris}, H.~V., {Komatsu}, E., {Nolta}, M.~R.,
  {Bennett}, C.~L., {Halpern}, M., {Hinshaw}, G., {Jarosik}, N., {Kogut}, A.,
  {Limon}, M., {Meyer}, S.~S., {Page}, L., {Tucker}, G.~S., {Weiland}, J.~L.,
  {Wollack}, E., \& {Wright}, E.~L. 2003, \apjs, 148, 175

\bibitem[{Sperhake {et~al.}(2005)Sperhake, Kelly, Laguna, Smith, \&
  Schnetter}]{SETAL05}
Sperhake, U., Kelly, B., Laguna, P., Smith, K.~L., \& Schnetter, E. 2005, Phys.
  Rev. D, 71, 123042

\bibitem[{Springel {et~al.}(2004)Springel, Di~Matteo, \& Hernquist}]{SDiMH04}
Springel, V., Di~Matteo, T., \& Hernquist, L. 2004, MNRAS, Submitted
  (astro-ph/0411108)

\bibitem[{Steed \& Weinberg(2003)}]{SW03}
Steed, A. \& Weinberg, D.~H. 2003, \apj Submitted (astro-ph/0311312)

\bibitem[{{Tegmark} {et~al.}(1997){Tegmark}, {Silk}, {Rees}, {Blanchard},
  {Abel}, \& {Palla}}]{tegmark_etal97}
{Tegmark}, M., {Silk}, J., {Rees}, M.~J., {Blanchard}, A., {Abel}, T., \&
  {Palla}, F. 1997, ApJ, 474, 1

\bibitem[{{Tegmark et al.}(2004)}]{tegmark_etal04}
{Tegmark et al.}, M. 2004, \prd, 69, 103501

\bibitem[{{Thorne}(1980)}]{thorne80}
{Thorne}, K.~S. 1980, Rev. Mod. Phys., 52, 299

\bibitem[{{Thorne}(1987)}]{thorne87}
{Thorne}, K.~S. 1987, in Three Hundred Years of Gravitation, 330

\bibitem[{{Thorne}(1995)}]{T95}
{Thorne}, K.~S. 1995, in Particle and Nuclear Astrophysics and Cosmology in the
  Next Millenium: Proc. 1994 Snowmass Summer Study, 109

\bibitem[{{Thorne}(1996)}]{thorne96}
{Thorne}, K.~S. 1996, in Compact Stars In Binaries, 151

\bibitem[{Thorne \& Braginsky(1976)}]{TB76}
Thorne, K.~S. \& Braginsky, V.~B. 1976, ApJ, 204, L1

\bibitem[{Tremaine {et~al.}(2002)}]{TETAL02}
Tremaine, S. {et~al.} 2002, ApJ, 574, 740

\bibitem[{{Turner}(1997)}]{turner97}
{Turner}, M.~S. 1997, PRD, 55, 435

\bibitem[{van~den Bosch {et~al.}(2002)van~den Bosch, Abel, Croft, Hernquist, \&
  White}]{VDBETAL02}
van~den Bosch, F.~C., Abel, T., Croft, R. A.~C., Hernquist, L., \& White, S.
  D.~M. 2002, ApJ, 576, 21

\bibitem[{Volonteri {et~al.}(2003)Volonteri, Haardt, \& Madau}]{VHM03}
Volonteri, M., Haardt, F., \& Madau, P. 2003, ApJ, 582, 559

\bibitem[{Volonteri {et~al.}(2005)Volonteri, Madau, Quataert, \& Rees}]{VMQR04}
Volonteri, M., Madau, P., Quataert, E., \& Rees, M.~J. 2005, Astrophys. J.,
  620, 69

\bibitem[{{White} \& {Rees}(1978)}]{white_rees78}
{White}, S.~D.~M. \& {Rees}, M.~J. 1978, MNRAS, 183, 341

\bibitem[{Wyithe \& Loeb(2003)}]{WL03}
Wyithe, J.~S. \& Loeb, A. 2003, ApJ, 590, 691

\bibitem[{Yu(2002)}]{Y02}
Yu, Q. 2002, MNRAS, 331, 935

\bibitem[{{Zaldarriaga} {et~al.}(2004){Zaldarriaga}, {Furlanetto}, \&
  {Hernquist}}]{cabron_etal04}
{Zaldarriaga}, M., {Furlanetto}, S.~R., \& {Hernquist}, L. 2004, \apj, 608, 622

\bibitem[{Zentner {et~al.}(2005)Zentner, Berlind, Bullock, Kravtsov, \&
  Wechsler}]{ZETAL04}
Zentner, A.~R., Berlind, A.~A., Bullock, J.~S., Kravtsov, A.~V., \& Wechsler,
  R.~H. 2005, ApJ, 624

\bibitem[{Zentner \& Bullock(2003)}]{ZB03}
Zentner, A.~R. \& Bullock, J.~S. 2003, ApJ, 598, 49

\end{thebibliography}

\end{document}